\documentclass[preprint,pre]{revtex4-1}%
\usepackage{amssymb}
\usepackage{amsfonts}
\usepackage{amsmath}
\usepackage{graphicx}
\usepackage{pgf}
\usepackage{tikz}
\usepackage{wrapfig}%
\providecommand{\U}[1]{\protect\rule{.1in}{.1in}}

\usepgflibrary{patterns}
\newcommand{\nodalna}[3]{
\draw[thick] (#1,#2) -> +(0,#3);}

\newcommand{\linprzer}[3]{
\draw[dashed,ultra thin] (#1,#2) -- +(#3,0);
}

\newcommand{\operM}[2]{
\filldraw[fill=white] (#1,#2) circle(.15);
}
\ifx\pdfoutput\relax\let\pdfoutput=\undefined\fi
\newcount\msipdfoutput
\ifx\pdfoutput\undefined\else
\ifcase\pdfoutput\else
\ifx\paperwidth\undefined\else
\ifdim\paperheight=0pt\relax\else\pdfpageheight\paperheight\fi
\ifdim\paperwidth=0pt\relax\else\pdfpagewidth\paperwidth\fi
\fi\fi\fi
\begin{document}
\preprint{HEP/123-qed}
\title[Generalization of Clausius-Mossotti approximation]{Generalization of Clausius-Mossotti approximation in application to short-time
transport properties of suspensions}
\author{Karol Makuch}
\email{Karol.Makuch@fuw.edu.pl}
\affiliation{Faculty of Physics, University of Warsaw}

\begin{abstract}
In 1983 Felderhof, Ford and Cohen gave microscopic explanation of the famous
Clausius-Mossotti formula for the dielectric constant of nonpolar dielectric.
They based their considerations on the cluster expansion of the dielectric
constant, which relates this macroscopic property with the microscopic
characteristics of the system.

In this article, we analyze the cluster expansion of Felderhof, Ford and Cohen
by performing its resummation (renormalization). Our analysis leads to the
ring expansion for the macroscopic characteristic of the system, which is an
expression alternative to the cluster expansion. Using similarity of
structures of the cluster expansion and the ring expansion, we generalize
(renormalize) the Clausius-Mossotti approximation. We apply our renormalized
Clausius-Mossotti approximation to the case of the short-time transport
properties of suspensions, calculating the effective viscosity and the
hydrodynamic function with the translational self-diffusion and the collective
diffusion coefficient. We
perform calculations for monodisperse hard-sphere suspensions in equilibrium
with volume fraction up to $45\%$. To assess the renormalized
Clausius-Mossotti approximation, it is compared with numerical simulations and
the Beenakker-Mazur method. The results of our renormalized Clausius-Mossotti
approximation lead to comparable or much less error (with respect to the
numerical simulations), than the Beenakker-Mazur method for the volume
fractions below $\phi\approx30\%$ (apart from a small range of wave vectors in hydrodynamic function).
For volume fractions above $\phi \approx30\%$,
the Beenakker-Mazur method gives in most cases lower error, than the renormalized Clausius-Mossotti approximation.

\end{abstract}
\volumeyear{year}
\volumenumber{number}
\issuenumber{number}
\eid{identifier}
\date[Date text]{date}
\received[Received text]{date}

\revised[Revised text]{date}

\accepted[Accepted text]{date}

\published[Published text]{date}

\maketitle

\section{Introduction}

Einstein was the first, who applied statistical physics to calculate the
viscosity of suspension \cite{einstein1906neue}. Having in mind nanometer size
sugar molecules immersed in water, he considered a model of sufficiently big
spherical particles immersed in viscous liquid. Experiments show, that in this
case the observed viscosity increases \cite{russel1992colloidal}. In his work,
Einstein related the observed (effective) viscosity $\eta_{\text{eff}}$ of
suspension with its structure on the microscopic level. His result,
$\eta_{\text{eff}}/\eta=1+5/2\phi$ - where $\eta$ denotes the viscosity of
solvent and $\phi$ denotes the volume fraction of the system - is valid only
for dilute suspensions. This limitation is caused by the assumption, that the
particles immersed in fluid do not influence their mutual motion. The problem
of the influence of the particles on their mutual motion is essential to go
beyond the diluted regime and was already addressed by Smoluchowski.
\begin{wrapfigure}{l}{5cm}%
\vspace{-20pt}
\includegraphics[
width=5.101cm
]
{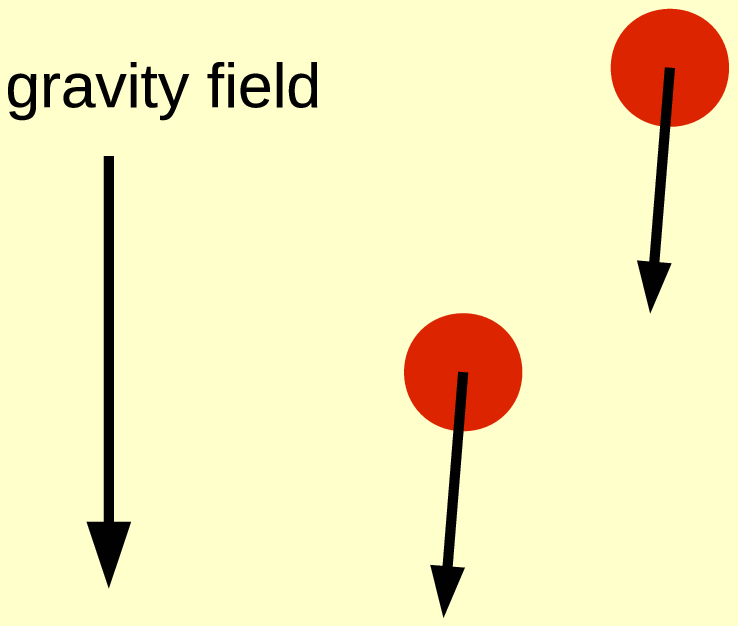}%
\vspace{-20pt}
\end{wrapfigure}His analysis for two sedimenting spheres leads to the
following conclusions. The two spheres sediment faster than a single one.
Moreover, the velocities of both spheres are deviated from the direction of
the gravity field, as shown in the figure. This example clearly demonstrates,
that two sedimenting particles in gravity field behave differently than a
single particle, because a single particle would sediment vertically downward.
Despite the fact, that there are no direct forces between the particles, they
influence their motion. This "interaction" of the immersed particles is
mediated by fluid and is called the "hydrodynamic interaction".

Apart from the considerations for the finite number of particles, Smoluchowski
analyzed also an infinite set of particles. He concluded, that behavior of
suspension strongly depends on the shape of the system. The shape matters,
even if its boundaries are extended to infinity. In other words, Smoluchowski
identified the problem of long-range hydrodynamic interactions. Another
important feature of the hydrodynamic interactions is their many-body
character. Motion of three particles cannot be described as a superposition of
the two-particle characteristics. Similar holds for larger number of
particles. In general, many-body characteristics are needed in the macroscopic
considerations for suspensions. From the perspective of transport properties,
even the two-body hydrodynamic interactions are problematic. Analysis of the
two-body problem reveals, that two particles at a small distance in
incompressible, viscous fluid, strongly "interact" hydrodynamically. In order
to keep constant velocity of the approaching particles, asymptotically an
infinite force is needed \cite{jeffrey1984calculation}.

Extension of Einstein's analysis for more concentrated systems appeared to be
difficult, because of the long-range hydrodynamic interactions. One of the
first successful approaches was made by Saito \cite{saito1950concentration},
who obtained the following formula for the effective viscosity, $\eta
_{\text{eff}}/\eta=\left(  1+3\phi/2\right)  /(1-\phi)$. Saito took the
hydrodynamic interactions into consideration partially. He also discussed the
long-range character of the hydrodynamic interactions and strongly emphasized
difficulties unsolved at that time \cite{saito1952remark}. The first systematic
extension of Einstein's work for more concentrated suspensions was performed
by Peterson and Fixman \cite{Peterson1963Viscosity}. They obtained a virial expansion
of the effective viscosity up to the second order, which includes the
two-body hydrodynamic interactions. It was the first approach, in which the
transport coefficient in the second order was given by absolutely convergent
integrals. Despite of this success, they did not express the transport
coefficients by absolutely convergent integrals for higher orders of virial
expansion. Therefore, the problem with long-range hydrodynamic interactions
was still not solved at that time. Solution came with the work of Felderhof,
Ford and Cohen in 1982 \cite{felderhof1982cluster}. The above authors
considered a dielectric system, but their analysis can be directly carried
over to the physics of suspensions. They proved, that the dielectric constant
is a local quantity, which does not depend on the shape of the system. Their
idea is related to the Brown's approach, who obtained similar result limited
to the lowest terms in the single-particle polarizability expansion of a
dielectric constant \cite{brown1955solid}. Felderhof, Ford and Cohen also gave
the microscopic explanation of the famous Clausius-Mossotti formula
\cite{felderhof1983clausius}, which is an analog of the Saito formula \cite{saito1950concentration} for the effective viscosity in the physics of
suspensions. It is worth mentioning here the effective medium approaches
\cite{mellema1983effective, bedeaux1983effective} and their extensions
including the two-body hydrodynamic interactions in a more accurate way
\cite{bedeaux1977effective, cichocki1989effective}.

Nowadays, the most prominent statistical physics approach to the short-time
transport properties of suspensions is the Beenakker-Mazur method
\cite{resummation83, Beenakker1984effective, Beenakker1984Diffusion},
which was developed and applied for different suspensions
\cite{genz1991collective,banchio2008short,westermeier2012structure,riest2015dynamics,heinen2011short,heinen2011short,Wang2015a}.
The method gives reasonable results for a wide range of volume fractions, but it
does not take the two-body hydrodynamic interactions fully into account. It is
known from virial expansion \cite{trojczastkowasamodyfuzja, Cichocki2002three,
trojczastkowalepkosc} and from numerical simulations,
\cite{abade:104902,durlofsky1987dynamic} that the two-body hydrodynamic
interactions of close particles are essential to grasp the dynamics of the
system. Therefore, there is still an open problem in the physics of
suspensions: formulation of a systematic method, which would take the two-body
hydrodynamic interactions fully into consideration and which would give
reasonable results for at least the intermediate volume fractions, say
$\phi\approx25\%$. Systematic consideration of the two-body hydrodynamic
interactions in the Beenakker-Mazur method is difficult, because the method
relies on the expansion of the transport properties in the series of the
so-called renormalized fluctuations. This series expansion is then truncated
in the second order in the fluctuations. To consider the full two-body
hydrodynamic interactions in the Beenakker-Mazur expansion, one needs
summation of all orders in the series, which is impossible in practice.

In this article, we develop the approach of Felderhof, Ford and Cohen. As
mentioned above, they introduced the cluster expansion of the macroscopic
characteristics of dispersive media such as e.g. the polarizability of
dielectric and the effective viscosity of suspension
\cite{felderhof1982cluster}. Felderhof, Ford and Cohen also gave the
microscopic explanation of the Clausius-Mossotti formula for dielectrics
(related to the Saito formula in case of suspensions)
\cite{felderhof1983clausius}. Their cluster expansion is a starting point of
this article. We perform a rigorous analysis of the cluster expansion leading
to a formula, which we call ring expansion of the macroscopic characteristics.
We also generalize the Clausius-Mossotti approximation, basing on a similarity
between the Felderhof, Ford and Cohen's cluster expansion and the ring
expansion introduced in this article. Using the generalized Clausius-Mossotti
approximation, we calculate the effective viscosity and the hydrodynamic function
(with the translational short-time self-diffusion and the collective diffusion
coefficient)  for suspension of monodisperse hard-spheres in equilibrium.

The generalization (renormalization) of the Clausius-Mossotti approximation based on the ring expansion
introduced in this article
is motivated by the results of the virial expansion for the effective viscosity
and the sedimentation coefficients \cite{trojczastkowalepkosc,Cichocki2002three}.
One of the dominant contributions to the virial expansion for the sedimentation coefficient
on the three-body level comes from the terms with a virtual overlap of spheres.
The idea of resummation of the above terms with a virtual overlap of spheres
for more dense suspensions was presented to the author of this article by Prof. Bogdan Cichocki,
to whom the author is very grateful.

The outline of this article is as follows. In second section, we describe the
suspension on the microscopic level and discuss the macroscopic
characteristics of suspensions. In third section, we repeat the analysis of
Felderhof, Ford and Cohen leading to the cluster expansion of the macroscopic
characteristics. In fourth section, we introduce the novel ring expansion of
the macroscopic characteristics, which is a rigorous result. The ring
expansion is further used in fifth section of the article, to introduce a
generalization of the Clausius-Mossotti approximation. Here, we also present
the short-time transport properties calculated by this novel method. The
generalized Clausius-Mossotti approximation is discussed and its results are
compared with the results of the numerical simulations
and with the Beenakker-Mazur method.

\section{Macroscopic properties of suspensions}

We consider suspension of hard spheres of radius $a$ in incompressible
Newtonian fluid of kinematic viscosity $\eta$. We also assume sufficiently
slow motion of the fluid and the condition of no slip on the surface of
immersed particles. As a result, the fluid is described by the stationary
Stokes equations with the stick boundary conditions
\cite{kim1991microhydrodynamics}. The stationary Stokes equations for the
problem of the suspension of $N$ spheres, centered at positions $X\equiv
\mathbf{R}_{1},\ldots,\mathbf{R}_{N},$ freely-moving in ambient flow
$\mathbf{v}_{0}\left(  \mathbf{r}\right)  ,$ under action of external force
density field $\mathbf{f}_{\text{ext}}\left(  \mathbf{r}\right)  ,$ can be
represented in the following integral form \cite{diag_ostateczna_wersja,
makuch2012scattering}%
\begin{align}
\mathbf{U}_{i}\left(  X,\mathbf{r}\right)   &  =\int d^{3}r^{\prime}%
\mathbf{M}_{0}\left(  \mathbf{R}_{i}\mathbf{,r,r}^{\prime}\right)
\mathbf{f}_{\text{ext}}\left(  \mathbf{r}^{\prime}\right)  \nonumber\\
&  +\int d^{3}r^{\prime}\mathbf{M}_{<}\left(  \mathbf{R}_{i}\mathbf{,r,r}%
^{\prime}\right)  \left[  \mathbf{v}_{0}\left(  \mathbf{r}^{\prime}\right)
+\sum_{\substack{j=1\\j\not =i}}^{N}\int d^{3}r^{\prime\prime}\mathbf{G}%
_{0}\left(  \mathbf{r}^{\prime},\mathbf{r}^{\prime\prime}\right)
\mathbf{f}_{j}\left(  X;\mathbf{r}^{\prime\prime}\right)  \right]
,\nonumber\\
\mathbf{f}_{i}\left(  X,\mathbf{r}\right)   &  =\int d^{3}r^{\prime
}\mathbf{\hat{M}}\left(  i\mathbf{,r,r}^{\prime}\right)  \left[
\mathbf{v}_{0}\left(  \mathbf{r}^{\prime}\right)  +\sum
_{\substack{j=1\\j\not =i}}^{N}\int d^{3}r^{\prime\prime}\mathbf{G}_{0}\left(
\mathbf{r}^{\prime},\mathbf{r}^{\prime\prime}\right)  \mathbf{f}_{j}\left(
X;\mathbf{r}^{\prime\prime}\right)  \right]  \nonumber\\
&  +\int d^{3}r^{\prime}\mathbf{M}_{>}\left(  i\mathbf{,r,r}^{\prime}\right)
\mathbf{f}_{\text{ext}}\left(  \mathbf{r}\right)  .\label{Stokes surface}%
\end{align}
In the above equations, the particle velocity field $\mathbf{U}_{i}\left(
X;\mathbf{r}\right)  $ is defined inside the particle, i.e. for $\left\vert
\mathbf{r-R}_{i}\right\vert \leq a$. For hard spheres it has always the
following form%
\begin{equation}
\mathbf{U}_{i}\left(  X;\mathbf{r}\right)  =\mathbf{V}_{i}\left(  X\right)
+\mathbf{\Omega}_{i}\left(  X\right)  \times\left(  \mathbf{r-R}_{i}\right)
,\ \ \ \ \ \ \text{for \ \ \ }\left\vert \mathbf{r-R}_{i}\right\vert \leq
a,\label{def particle field}%
\end{equation}
with translational $\mathbf{V}_{i}$ and rotational $\mathbf{\Omega}_{i}$
velocity of the particles. Moreover, $\mathbf{f}_{i}\left(  X;\mathbf{r}%
\right)  $ describes the force density \cite{Mazur1974235, felderhof1988many,
marysiaElekFragment} acting on the fluid by the surface of the particle number
$i$ and is defined by%
\begin{equation}
\mathbf{f}_{i}\left(  X;\mathbf{r}\right)  =-\mathbf{\sigma}\left(
\mathbf{r}\right)  \cdot\mathbf{\hat{n}}_{i}\left(  \mathbf{r}\right)
\mathbf{\ }\delta\left(  \left\vert \mathbf{r-R}_{i}\right\vert -a\right)
,\label{sila indukowana a tensor cisnien}%
\end{equation}
where $\mathbf{\sigma}$ represents the stress tensor in the fluid,
$\mathbf{\hat{n}}_{i}\left(  \mathbf{r}\right)  =\left(  \mathbf{r-R}%
_{i}\right)  /\left\vert \mathbf{r-R}_{i}\right\vert $ is a vector normal to
the surface of the sphere $i$, whereas $\delta\left(  x\right)  $ stands for
the one-dimensional Dirac delta function. $\mathbf{G}_{0}\left(
\mathbf{r}\right)  $ in equations (\ref{Stokes surface}) is the Oseen tensor,%
\begin{equation}
\mathbf{G}_{0}\left(  \mathbf{r}\right)  =\left(  \mathbf{1}+\mathbf{\hat
{r}\hat{r}}\right)  /\left(  8\pi\eta\left\vert \mathbf{r}\right\vert \right)
,
\end{equation}
with $\mathbf{\hat{r}}=\mathbf{r/}\left\vert \mathbf{r}\right\vert $. The
Oseen tensor is a Green function of the Stokes equations
\cite{ladyzhenskaya:57}, hence flow of the whole suspension $\mathbf{v}\left(
\mathbf{r}\right)  $ is given by
\begin{equation}
\mathbf{v}\left(  \mathbf{r}\right)  =\mathbf{v}_{0}\left(  \mathbf{r}\right)
+\sum_{i=1}^{N}\int d^{3}r^{\prime}\mathbf{G}_{0}\left(  \mathbf{r}%
-\mathbf{r}^{\prime}\right)  \mathbf{\cdot f}_{i}\left(  \mathbf{r}^{\prime
}\right)  .\label{flow in suspension}%
\end{equation}
The equations (\ref{Stokes surface}) are linear both in the ambient flow
$\mathbf{v}_{0}$ and in the external force density $\mathbf{f}_{\text{ext}}.$
Therefore, to describe the response operators $\mathbf{M}_{0}$, $\mathbf{M}%
_{<}$, $\mathbf{\hat{M}}$, and $\mathbf{M}_{>},$ it is sufficient and
convenient to consider special cases of a single particle problem.
$\mathbf{M}_{0}$ in the equations (\ref{Stokes surface}), in the case of the
single particle problem in the external force density field $\mathbf{f}%
_{\text{ext}}\left(  \mathbf{r}\right)  ,$ and in absence of the ambient flow,
$\mathbf{v}_{0}=0$, yields the velocity field of the particle,%
\begin{equation}
\mathbf{U}_{1}\left(  \mathbf{R}_{1};\mathbf{r}\right)  =\int d^{3}r^{\prime}
\mathbf{M}_{0}\left(  \mathbf{R}_{1}\mathbf{,r,r}^{\prime}\right)
\mathbf{f}_{\text{ext}}\left(  \mathbf{r}^{\prime}\right)  .
\end{equation}
The single particle operator $\mathbf{M}_{<}$ gives the particle velocity
field $\mathbf{U}_{1},$ when the particle is placed in the ambient flow
$\mathbf{v}_{0}$,%
\begin{equation}
	\mathbf{U}_{1}\left(  \mathbf{R}_{1};\mathbf{r}\right)  =\int d^{3}r^{\prime}
\mathbf{M}_{<}\left(  \mathbf{R}_{1}\mathbf{,r,r}^{\prime}\right)
\mathbf{v}_{0}\left(  \mathbf{r}^{\prime}\right)  .
\end{equation}
$\mathbf{\hat{M}}\left(  \mathbf{R,r,r}^{\prime}\right)  $, called the single
particle convective friction kernel, yields the force density $\mathbf{f}%
_{1}\left(  \mathbf{R}_{1};\mathbf{r}\right)  $ on the surface of the single
particle at the position $\mathbf{R}_{1},$ when it is placed in the ambient
flow $\mathbf{v}_{0}\left(  \mathbf{r}\right)  $,%
\begin{equation}
\mathbf{f}_{1}\left(  \mathbf{R}_{1};\mathbf{r}\right)  =\int d^{3}r^{\prime
}\mathbf{\hat{M}}\left(  \mathbf{R}_{1}\mathbf{,r,r}^{\prime}\right)
\mathbf{v}_{0}\left(  \mathbf{r}^{\prime}\right)  .
\end{equation}
Finally, $\mathbf{M}_{>}$ describes the force density $\mathbf{f}_{1}\left(
\mathbf{R}_{1};\mathbf{r}\right)  $ on the surface of the single particle at
the position $\mathbf{R}_{1},$ under the action of the external force
$\mathbf{f}_{\text{ext}}$,%
\begin{equation}
\mathbf{f}_{1}\left(  \mathbf{R}_{1};\mathbf{r}\right)  =\int d^{3}r^{\prime
}\mathbf{M}_{>}\left(  \mathbf{R}_{1}\mathbf{,r,r}^{\prime}\right)
\mathbf{f}_{\text{ext}}\left(  \mathbf{r}^{\prime}\right)  .\label{single M>}%
\end{equation}
In this article, we investigate the equations (\ref{Stokes surface}) mostly
without referring to the specific form of the response operators
$\mathbf{M}_{0}$, $\mathbf{M}_{<}$, $\mathbf{M}_{>}$, $\mathbf{\hat{M}}$. For
their detail form, we refer the reader to the references
\cite{makuch2012scattering, Felderhof1976force2}.

To facilitate further analysis of the equations (\ref{Stokes surface}), we
omit integral variables in those equations, writing them in the following form%
\begin{align}
\mathbf{U}_{i}\left(  X\right)   &  =\mathbf{M}_{0}\left(  i\right)
\mathbf{f}_{\text{ext}}+\mathbf{M}_{<}\left(  i\right)  \left[  \mathbf{v}%
_{0}+\sum_{\substack{j=1\\j\not =i}}^{N}\mathbf{G}_{0}\mathbf{f}_{j}\left(
X\right)  \right]  ,\nonumber\\
\mathbf{f}_{i}\left(  X\right)   &  =\mathbf{\hat{M}}\left(  i\right)  \left[
\mathbf{v}_{0}+\sum_{\substack{j=1\\j\not =i}}^{N}\mathbf{G}_{0}\mathbf{f}%
_{j}\left(  X\right)  \right]  +\mathbf{M}_{>}\left(  i\right)  \mathbf{f}%
_{\text{ext}}. \label{Stokes mobility integral form}%
\end{align}
For the position of the particle $i$ in the single particle response operators
$\mathbf{M}_{0}$, $\mathbf{M}_{<}$, $\mathbf{\hat{M}}$, $\mathbf{M}_{>},$ we
also use the following abbreviation: $i\equiv\mathbf{R}_{i}$. Finally, we
write the above equations as follows \cite{makuch2012scattering},
\begin{equation}
\left[
\begin{array}
[c]{c}%
\mathbf{U}_{i}\left(  X\right) \\
\mathbf{f}_{i}\left(  X\right)
\end{array}
\right]  =\mathbf{M}\left(  i\right)  \left(  \left[
\begin{array}
[c]{c}%
\mathbf{f}_{\text{ext}}\\
\mathbf{v}_{0}%
\end{array}
\right]  +\sum_{\substack{j=1\\j\not =i}}^{N}\mathbf{G}\left[
\begin{array}
[c]{c}%
\mathbf{U}_{j}\left(  X\right) \\
\mathbf{f}_{j}\left(  X\right)
\end{array}
\right]  \right)  , \label{Stokes mobility integral forrm}%
\end{equation}
introducing $6\times6$ dimensional matrices $\mathbf{M}$ and $\mathbf{G}$
defined by the below equations,%
\begin{equation}
\mathbf{M}\left(  \mathbf{R,r,r}^{\prime}\right)  =\left[
\begin{array}
[c]{cc}%
\mathbf{M}_{0}\left(  \mathbf{R,r,r}^{\prime}\right)  & \mathbf{M}_{<}\left(
\mathbf{R,r,r}^{\prime}\right) \\
\mathbf{M}_{>}\left(  \mathbf{R,r,r}^{\prime}\right)  & \mathbf{\hat{M}%
}\left(  \mathbf{R,r,r}^{\prime}\right)
\end{array}
\right]
\end{equation}
and%
\begin{equation}
\mathbf{G}\left(  \mathbf{r},\mathbf{r}^{\prime}\right)  \mathbf{=}\left[
\begin{array}
[c]{cc}%
\mathbf{0} & \mathbf{0}\\
\mathbf{0} & \mathbf{G}_{0}\left(  \mathbf{r},\mathbf{r}^{\prime}\right)
\end{array}
\right]  .
\end{equation}

\subsection{Scattering series}

To solve the equations (\ref{Stokes mobility integral forrm}), several methods
can be used. One of the possible approaches is the method of reflections
\cite{smoluchowski1912practical}. It relies on taking successive iterations of
the equation (\ref{Stokes mobility integral forrm}) which leads to the
following formula:%
\begin{align}
\left[
\begin{array}
[c]{c}%
\mathbf{U}_{i}\left(  X\right) \\
\mathbf{f}_{i}\left(  X\right)
\end{array}
\right]   &  =\mathbf{M}\left(  i\right)  \left[
\begin{array}
[c]{c}%
\mathbf{f}_{\text{ext}}\\
\mathbf{v}_{0}%
\end{array}
\right]  +\sum_{\substack{j=1,\\j\not =i}}^{N}\mathbf{M}\left(  i\right)  \mathbf{GM}\left(
j\right)  \left[
\begin{array}
[c]{c}%
\mathbf{f}_{\text{ext}}\\
\mathbf{v}_{0}%
\end{array}
\right] \nonumber\\
&  +\sum_{\substack{j=1,\\j\not =i}}^{N}\sum_{\substack{k=1,\\k\not =j}%
}^{N}\mathbf{M}\left(  i\right)  \mathbf{GM}\left(  j\right)  \mathbf{GM}%
\left(  k\right)  \left[
\begin{array}
[c]{c}%
\mathbf{f}_{\text{ext}}\\
\mathbf{v}_{0}%
\end{array}
\right]  +\ldots. \label{si przez sekwencje rozproszeniowe}%
\end{align}
The above expression of the force densities, $\mathbf{f}_{i}\left(  X\right)
,$ and the velocities of the particles, $\mathbf{U}_{i}\left(  X\right)  ,$
has a form of a multiple scattering series. It means, that $\mathbf{f}%
_{i}\left(  X\right)  $ and $\mathbf{U}_{i}\left(  X\right)  $ are given by
the sum of the scattering sequences, for example:%
\begin{equation}
\mathbf{M}\left(  1\right)  \mathbf{GM}\left(  2\right)  \label{scat seq ex1}%
\end{equation}
and%
\begin{equation}
\mathbf{M}\left(  1\right)  \mathbf{GM}\left(  2\right)  \mathbf{GM}\left(
3\right)  \mathbf{GM}\left(  2\right)  . \label{scat seq ex2}%
\end{equation}
As we interpret - each scattering sequence is a superposition of the
single-particle scattering operators $\mathbf{M}\left(  i\right)  ,$ which
"scatter" the flow and of Green functions $\mathbf{G,}$ which "propagates" the flow.

It is convenient and useful to represent the scattering sequences graphically
\cite{diag_ostateczna_wersja}. The above two sequences can be represented
respectively by
\begin{equation}
\text{\parbox[c]{2.1cm}{
\begin{tikzpicture}
\linprzer{-0.1}{0}{1.3} \linprzer{-0.1}{1}{1.3}
\nodalna{0.45}{0}{1}
\operM{.3}{0}  \operM{.6}{1}
\draw (-0.1,0) node[anchor=east] {1};
\draw (-0.1,1) node[anchor=east] {2};
\end{tikzpicture}
},\parbox[c]{3.1cm}{
\begin{tikzpicture}
\linprzer{-0.2}{0}{1.9} \linprzer{-0.2}{1}{1.9} \linprzer{-0.2}{2}{1.9}
\nodalna{0.35}{0}{1} \nodalna{.65}{1}{1} \nodalna{.95}{1}{1}
\operM{.2}{0}  \operM{.5}{1}  \operM{0.8}{2}  \operM{1.1}{1}
\draw(-0.2,0) node[anchor=east] {1};
\draw(-0.2,1) node[anchor=east] {2};
\draw(-0.2,2) node[anchor=east] {3};
\end{tikzpicture}
}.}%
\end{equation}
In general, to represent a scattering sequence graphically, we draw horizontal
dashed lines \parbox[c]{1.1cm}{
\begin{tikzpicture}
\linprzer{0}{0}{1}
\end{tikzpicture}
}. Each line corresponds to a particle in the scattering sequence. Then,
reading the sequence from left to right, we put successively: the circle
\parbox[c]{0.6cm}{
\begin{tikzpicture}
\operM{0}{0}
\end{tikzpicture}
} on the dashed line $i$ for the operator $\mathbf{M}\left(  i\right)  $ and
the vertical line \parbox[c]{0.2cm}{
\begin{tikzpicture}
\nodalna{0}{0}{0.5}
\end{tikzpicture}
} connecting the dashed lines $i$ and $j$ for the Oseen tensor $\mathbf{G,}$
when it appears in the configuration $\mathbf{M}\left(  i\right)
\mathbf{GM}\left(  j\right)  $.

The scattering series plays a major role in our considerations. We denote the
scattering series by $\mathbf{T}_{ij}\left(  X\right)  $:%
\begin{align}
\mathbf{T}_{ij}\left(  X\right)   &  =\mathbf{M}\left(  i\right)  \delta
_{ij}+\mathbf{M}\left(  i\right)  \mathbf{GM}\left(  j\right)  \left(
1-\delta_{ij}\right) \nonumber\\
&  +\sum_{\substack{k=1\\k\neq i,k\neq j}}^{N}\mathbf{M}\left(  i\right)
\mathbf{GM}\left(  k\right)  \mathbf{GM}\left(  j\right)  +\ldots.
\label{scattering series}%
\end{align}
Therefore, the velocity $\mathbf{U}_{i}\left(  X\right)  $ and the force
density $\mathbf{f}_{i}\left(  X\right)  $ in the expression
(\ref{si przez sekwencje rozproszeniowe}) are given by the formula%

\begin{equation}
\left[
\begin{array}
[c]{c}%
\mathbf{U}_{i}\left(  X\right) \\
\mathbf{f}_{i}\left(  X\right)
\end{array}
\right]  =\sum_{j=1}^{N}\mathbf{T}_{ij}\left(  X\right)  \left[
\begin{array}
[c]{c}%
\mathbf{f}_{\text{ext}}\\
\mathbf{v}_{0}%
\end{array}
\right]  . \label{f by Tij and v0}%
\end{equation}

\subsection{Macroscopic response}

To describe properties of suspension on the macroscopic level, we consider an
ensemble of configurations of particles $X\equiv\mathbf{R}_{1},\ldots
,\mathbf{R}_{N},$ which is described by a probability distribution function
$p\left(  X\right)  $. We also introduce the average force density defined by
the equation%
\begin{equation}
\left\langle \mathbf{f}\left(  \mathbf{R},\mathbf{r}\right)  \right\rangle
=\left\langle \sum_{i=1}^{N}\delta\left(  \mathbf{R}-i\right)  \mathbf{f}%
_{i}\left(  X,\mathbf{r}\right)  \right\rangle
\end{equation}
and the average particle velocity field%
\begin{equation}
\left\langle \mathbf{U}\left(  \mathbf{R},\mathbf{r}\right)  \right\rangle
=\left\langle \sum_{i=1}^{N}\delta\left(  \mathbf{R}-i\right)  \mathbf{U}%
_{i}\left(  X,\mathbf{r}\right)  \right\rangle ,
\end{equation}
where the three-dimensional Dirac delta function $\delta\left(  \mathbf{R}%
-i\right)  \equiv\delta\left(  \mathbf{R}-\mathbf{R}_{i}\right)  $ and the
average over the probability distribution $\left\langle \left[  \ldots\right]
\right\rangle =\int d^{3}R_{1}\ldots\int d^{3}R_{N}\ p\left(  X\right)
\left[  \ldots\right]  $ are used. Averages of the equations
(\ref{f by Tij and v0}) - multiplied by proper Dirac delta functions - lead to
the following expression for the average velocity and the average force
density,%
\begin{equation}
\left[
\begin{array}
[c]{c}%
\left\langle \mathbf{U}\left(  \mathbf{R},\mathbf{r}\right)  \right\rangle \\
\left\langle \mathbf{f}\left(  \mathbf{R},\mathbf{r}\right)  \right\rangle
\end{array}
\right]  =\int d^{3}R^{\mathbf{\prime}}d^{3}r^{\mathbf{\prime}}\mathbf{T}%
\left(  \mathbf{R,r};\mathbf{R}^{\prime},\mathbf{r}^{\prime}\right)  \left[
\begin{array}
[c]{c}%
\mathbf{f}_{\text{ext}}\left(  \mathbf{r}^{\prime}\right) \\
\mathbf{v}_{0}\left(  \mathbf{r}^{\prime}\right)
\end{array}
\right]  , \label{def T}%
\end{equation}
where the averaged scattering series is denoted by $\mathbf{T}\left(
\mathbf{R,r};\mathbf{R}^{\prime},\mathbf{r}^{\prime}\right)  $ and defined
with the formula%
\begin{equation}
\mathbf{T}\left(  \mathbf{R,r};\mathbf{R}^{\prime},\mathbf{r}^{\prime}\right)
=\left\langle \sum_{i=1}^{N}\sum_{j=1}^{N}\delta\left(  \mathbf{R}-i\right)
\mathbf{T}_{ij}\left(  X,\mathbf{r},\mathbf{r}^{\prime}\right)  \delta\left(
\mathbf{R}^{\prime}-j\right)  \right\rangle . \label{T micro}%
\end{equation}
Notice, that in the above operator $\mathbf{T}\left(  \mathbf{R,r}%
;\mathbf{R}^{\prime},\mathbf{r}^{\prime}\right)  ,$ the Dirac delta functions
fix positions of the first ($i$) and the last ($j$) particle in the scattering
series $\mathbf{T}_{ij}$ at the positions $\mathbf{R}$ and $\mathbf{R}%
^{\prime}$ respectively. The average flow of the suspension $\left\langle
\mathbf{v}\left(  \mathbf{r}\right)  \right\rangle $ is a combination of the
ambient flow $\mathbf{v}_{0}\left(  \mathbf{r}\right)  ,$ in which the
particles are immersed and of flow generated by the presence of the particles%
\begin{equation}
\left\langle \mathbf{v}\left(  \mathbf{r}\right)  \right\rangle =\mathbf{v}%
_{0}\left(  \mathbf{r}\right)  +\int d^{3}R\int d^{3}r^{\prime}\mathbf{G}%
_{0}\left(  \mathbf{r},\mathbf{r}^{\prime}\right)  \left\langle \mathbf{f}%
\left(  \mathbf{R},\mathbf{r}^{\prime}\right)  \right\rangle ,
\label{av velocity field}%
\end{equation}
which is obtained by averaging the formula (\ref{flow in suspension}). We
eliminate the flow $\mathbf{v}_{0}$ from the equations
(\ref{av velocity field}) and (\ref{def T}), which leads to the formula%

\begin{equation}
\left[
\begin{array}
[c]{c}%
\left\langle \mathbf{U}\right\rangle \\
\left\langle \mathbf{f}\right\rangle
\end{array}
\right]  =\mathbf{T}\left[
\begin{array}
[c]{c}%
\mathbf{f}_{\text{ext}}\\
\left\langle \mathbf{v}\right\rangle
\end{array}
\right]  -\mathbf{TG}\left[
\begin{array}
[c]{c}%
\left\langle \mathbf{U}\right\rangle \\
\left\langle \mathbf{f}\right\rangle
\end{array}
\right]  ,
\end{equation}
in which we also facilitate the notation by omitting the integral variables.
Its subsequent iterations lead to a relation of the average particle velocity
$\left\langle \mathbf{U}\right\rangle $ and the force density $\left\langle
\mathbf{f}\right\rangle $ to the external force density $\mathbf{f}%
_{\text{ext}}$ and the average flow of suspension $\left\langle \mathbf{v}%
\right\rangle ,$%
\begin{equation}
\left[
\begin{array}
[c]{c}%
\left\langle \mathbf{U}\left(  \mathbf{R,r}\right)  \right\rangle \\
\left\langle \mathbf{f}\left(  \mathbf{R,r}\right)  \right\rangle
\end{array}
\right]  =\int d^{3}R^{\mathbf{\prime}}d^{3}r^{\mathbf{\prime}}\mathbf{T}%
^{\text{irr}}\left(  \mathbf{R,r};\mathbf{R}^{\prime},\mathbf{r}^{\prime
}\right)  \left[
\begin{array}
[c]{c}%
\mathbf{f}_{\text{ext}}\left(  \mathbf{r}^{\prime}\right) \\
\left\langle \mathbf{v}\left(  \mathbf{r}^{\prime}\right)  \right\rangle
\end{array}
\right]  , \label{def Tirr}%
\end{equation}
which defines $\mathbf{T}^{\text{irr}}$ operator given by%
\begin{equation}
\mathbf{T}^{\text{irr}}=\mathbf{T}\left(  1+\mathbf{GT}\right)  ^{-1}.
\label{Tirr and T}%
\end{equation}

The equation (\ref{def Tirr}) is directly related to the macroscopic
properties of the suspension. For example, the effective viscosity
$\eta_{\text{eff}}$ can be inferred from the relation between the average
force density $\left\langle \mathbf{f}\left(  \mathbf{R,r}\right)
\right\rangle $ and the average suspension flow $\left\langle \mathbf{v}%
\left(  \mathbf{r}\right)  \right\rangle ,$ when no external forces act on the
particles, $\mathbf{f}_{\text{ext}}=0$. The relation between $\left\langle
\mathbf{f}\left(  \mathbf{R,r}\right)  \right\rangle $ and $\left\langle
\mathbf{v}\left(  \mathbf{r}\right)  \right\rangle $ in this situation results
from the equation (\ref{def Tirr}), after projection it into the lower half of
the double vectors $\left[  \left\langle \mathbf{U}\right\rangle ,\left\langle
\mathbf{f}\right\rangle \right]  $ and $\left[  \mathbf{f}_{\text{ext}%
},\left\langle \mathbf{v}\right\rangle \right]  $. In order to do that, we
introduce a projector $P_{L}$ defined by%
\begin{equation}
P_{L}\left[
\begin{array}
[c]{c}%
\left\langle \mathbf{U}\right\rangle \\
\left\langle \mathbf{f}\right\rangle
\end{array}
\right]  =\left\langle \mathbf{f}\right\rangle , \label{def Pl}%
\end{equation}
with its transposition $P_{L}^{T}$. After projection, the equation
(\ref{def Tirr}) reads%
\begin{equation}
\left\langle \mathbf{f}\left(  \mathbf{R,r}\right)  \right\rangle =\int
d^{3}R^{\mathbf{\prime}}d^{3}r^{\mathbf{\prime}}P_{L}\mathbf{T}^{\text{irr}%
}\left(  \mathbf{R,r};\mathbf{R}^{\prime},\mathbf{r}^{\prime}\right)
P_{L}^{T}\left\langle \mathbf{v}\left(  \mathbf{r}^{\prime}\right)
\right\rangle .
\end{equation}
If the $\mathbf{T}^{irr}$ operator is known, by calculating the following four
rank Cartesian tensor%
\begin{equation}
X_{\alpha\beta\delta\gamma}\left(  \mathbf{R},\mathbf{R}^{\prime}\right)
=\int d^{3}r\int d^{3}r^{\prime}\left(  \mathbf{r-R}\right)  _{\alpha}\left[
P_{L}\mathbf{T}^{\text{irr}}\left(  \mathbf{R,r};\mathbf{R}^{\prime
},\mathbf{r}^{\prime}\right)  P_{L}^{T}\right]  _{\beta\delta}\left(
\mathbf{r}^{\prime}\mathbf{-R}^{\prime}\right)  _{\gamma}, \label{eta micro 1}%
\end{equation}
and by symmetrizing it over the first and the second pair of the Cartesian
indexes%
\begin{equation}
X_{\alpha\beta\delta\gamma}^{dd}\left(  \mathbf{R},\mathbf{R}^{\prime}\right)
=\frac{1}{4}\left(  X_{\alpha\beta\delta\gamma}\left(  \mathbf{R}%
,\mathbf{R}^{\prime}\right)  +X_{\beta\alpha\delta\gamma}\left(
\mathbf{R},\mathbf{R}^{\prime}\right)  +X_{\alpha\beta\gamma\delta}\left(
\mathbf{R},\mathbf{R}^{\prime}\right)  +X_{\beta\alpha\gamma\delta}\left(
\mathbf{R},\mathbf{R}^{\prime}\right)  \right)  ,
\end{equation}
we obtain the effective viscosity $\eta_{\text{eff}},$ using the formula
\cite{diag_ostateczna_wersja, kim1991microhydrodynamics}
\begin{equation}
	\eta_{\text{eff}}=\eta+\lim_{\infty}\frac{1}{10N}\sum_{\alpha,\beta=1}^3\int d^{3}R\int
d^{3}R^{\prime}X_{\alpha\beta\beta\alpha}^{dd}\left(  \mathbf{R}%
,\mathbf{R}^{\prime}\right)  . \label{eta micro 3}%
\end{equation}
Thermodynamic limit $\lim_{\infty}$ is performed in the above equation.

Apart from the short-time effective viscosity $\eta_{\text{eff}},$ we also
consider the short-time wave dependent sedimentation coefficient $H\left(
q\right)  $. The sedimentation coefficient describes response of the
suspension to the external force of the plane wave form,%
\begin{equation}
\mathbf{F}_{\text{ext}}\left(  \mathbf{R}\right)  =F_{0}\mathbf{\hat
{q}\operatorname{Re}}\exp\left(  -i\mathbf{qR}\right)  .
\end{equation}
We show in the appendix \ref{app hq}, that under the action of the above
force, the average translational velocity of the particles defined by%
\begin{equation}
\left\langle \mathbf{V}\left(  \mathbf{R}\right)  \right\rangle =\left\langle
\sum_{i=1}^{N}\delta\left(  \mathbf{R-R}_{i}\right)  \mathbf{V}_{i}\left(
X\right)  \right\rangle ,
\end{equation}
in an isotropic and homogeneous suspension, has also a plane wave form,
\begin{equation}
\left\langle \mathbf{V}\left(  \mathbf{R}\right)  \right\rangle =V\left(
q\right)  \mathbf{\hat{q}\operatorname{Re}}\exp\left(  -i\mathbf{qR}\right)  .
\end{equation}
Linearity of the Stokes equations implies, that the coefficient $V\left(
q\right)  $ in the above formula is proportional to the force $F_{0}$,%
\[
V\left(  q\right)  =H\left(  q\right)  \mu_{0}F_{0}.
\]
This formula defines the wave dependent sedimentation coefficient $H\left(
q\right)  ,$ which is also called the hydrodynamic function. The factor
$\mu_{0}=1/(6\pi\eta a)$ denotes the Stokes coefficient. $H\left(  q\right)  $
is a dimensionless function with the property $H\left(  q\right)
\rightarrow1$ in the limit of a diluted suspension, i.e. when the volume
fraction $\phi\rightarrow0$. As we also discuss in the appendix \ref{app hq},
the microscopic expression for the hydrodynamic function $H\left(  q\right)  $
has the following form \cite{diag_ostateczna_wersja}%
\begin{equation}
H\left(  q\right)  =\frac{1}{\mu_{0}}\frac{1}{3}\text{Tr}\left[  \int
d^{3}R\ e^{-i\mathbf{q\cdot R}}Y\left(  \mathbf{R}\right)  \right]  ,
\label{hyd fun  mikro}%
\end{equation}
where $3\times3$ matrix $Y\left(  \mathbf{R}\right)  $ is defined by the
following equation%
\begin{equation}
Y\left(  \mathbf{R}-\mathbf{R}^{\prime}\right)  =\frac{1}{\left(  \frac{4}%
{3}\pi a^{3}\right)  ^{2}}\lim_{\infty}\int d^{3}r\int d^{3}r^{\prime}%
P_{U}\mathbf{T}^{\text{irr}}\left(  \mathbf{R,r};\mathbf{R}^{\prime
},\mathbf{r}^{\prime}\right)  P_{U}^{T}. \label{hyd fun micro 2}%
\end{equation}
The projector $P_{U}$ projects on the upper half of the double vectors
$\left[  \left\langle \mathbf{U}\right\rangle ,\left\langle \mathbf{f}%
\right\rangle \right]  ,$%
\begin{equation}
P^{U}\left[
\begin{array}
[c]{c}%
\left\langle \mathbf{U}\right\rangle \\
\left\langle \mathbf{f}\right\rangle
\end{array}
\right]  =\left\langle \mathbf{U}\right\rangle . \label{def Pu}%
\end{equation}
$P_{U}^{T}$ denotes transposition of $P_{U}.$

It is worth noting, that the hydrodynamic function for the zero wave vector,
$q=0$, describes the sedimentation rate, $K$, of the suspension in a gravity field,
\begin{equation}
K=H\left(  q=0\right)	\label{sedimentation}
\end{equation}
and is also related to the short-time collective diffusion coefficient $D_{c}$,%
\begin{equation}
D_{c}=D_{0}H\left(  q=0\right)  , \label{collective dif}%
\end{equation}
whereas for infinite wave vector length $H\left(  q\rightarrow\infty\right)  $
is related to the short-time self-diffusion coefficient $D_{s}$,%
\begin{equation}
D_{s}=D_{0}H\left(  q\rightarrow\infty\right)  . \label{self dif}%
\end{equation}
In both expressions $D_{0}=k_{B}T/\left(  6\pi\eta a\right)  $ is the
diffusion coefficient of a single particle.

Both, the effective viscosity, and the hydrodynamic function can be inferred
from the $\mathbf{T}^{\text{irr}}$ operator. It is shown by the expressions
(\ref{eta micro 1}-\ref{eta micro 3}) for the effective viscosity
$\eta_{\text{eff}}$ and by the equations (\ref{hyd fun mikro}%
-\ref{hyd fun micro 2}) for the wave dependent sedimentation coefficient
$H\left(  q\right)  $. Therefore, $\mathbf{T}^{\text{irr}}$ becomes the
quantity of the main interest in this article.

\section{Felderhof, Ford and Cohen analysis of $\mathbf{T}^{\text{irr}}$}

In the first stage of our analysis of $\mathbf{T}^{\text{irr}}$ defined by the
equation (\ref{def Tirr}), we follow the idea of Felderhof, Ford and Cohen.
They obtained the microscopic expression for $\mathbf{T}^{\text{irr}}$ for the
dielectric system in the form of a cluster expansion
\cite{felderhof1982cluster}. The application of their idea to the physics of
suspensions is straightforward, because the governing equations are similar
for suspensions and dielectrics \cite{batchelor1974transport,
diag_ostateczna_wersja}. To perform the cluster expansion of the operator
$\mathbf{T}^{\text{irr}}$on the basis of the expression (\ref{Tirr and T}),
Felderhof, Ford and Cohen introduced the cluster expansion of the operator
$\mathbf{T}$.

\subsection{Cluster expansion of $\mathbf{T}$}

In the expression (\ref{T micro}), $\mathbf{T}_{ij}\left(  \mathbf{r}%
,\mathbf{r}^{\prime};X\right)  $ includes infinitely many scattering
sequences, as shown in the formula (\ref{scattering series}). There are
scattering sequences with different number of particles: single particle
scattering sequences, e.g.%
\begin{equation}
\text{\parbox[c]{1.5cm}{
\begin{tikzpicture}
\linprzer{-0.1}{0}{0.8}
\operM{.3}{0}
\draw (-0.1,0) node[anchor=east] {1};
\end{tikzpicture}
},\parbox[c]{1.5cm}{
\begin{tikzpicture}
\linprzer{-0.1}{0}{0.8}
\operM{.3}{0}
\draw (-0.1,0) node[anchor=east] {4};
\end{tikzpicture}
},}%
\end{equation}
two-particle scattering sequences, e.g.%
\begin{equation}
\text{%
\parbox[c]{1.8cm}{
\begin{tikzpicture}[scale=0.8]
\linprzer{0}{0}{1.4} \linprzer{0}{1}{1.4}
\nodalna{0.35}{0}{1} \nodalna{.65}{0}{1} \nodalna{.95}{0}{1}
\operM{.2}{0}  \operM{.5}{1}  \operM{0.8}{0}  \operM{1.1}{1}
\draw(0,0) node[anchor=east] {1};
\draw(0,1) node[anchor=east] {2};
\end{tikzpicture}
}%
,\ \
\parbox[c]{2cm}{
\begin{tikzpicture}[scale=0.8]
\linprzer{0}{0}{1.7} \linprzer{0}{1}{1.7}
\nodalna{0.35}{0}{1} \nodalna{.65}{0}{1} \nodalna{.95}{0}{1} \nodalna{1.25}%
{0}{1}
\operM{.2}{0}  \operM{.5}{1}  \operM{0.8}{0}  \operM{1.1}{1}  \operM{1.4}{0}
\draw(0,0) node[anchor=east] {2};
\draw(0,1) node[anchor=east] {4};
\end{tikzpicture}
}%
,}%
\end{equation}
and scattering sequences with higher number of particles, up to $N$. The
scattering sequences with the same number of particles may include different
particles. It is noticeable in the examples above, where the first scattering
sequence is between the particles from the group $C=\left\{  1,2\right\}  $.
The second scattering sequence is between the particles from the group
$C=\left\{  2,4\right\}  $. All the scattering sequences $\sum_{i=1}^{N}%
\sum_{j=1}^{N}\delta\left(  \mathbf{R}-i\right)  \mathbf{T}_{ij}\left(
\mathbf{r},\mathbf{r}^{\prime};X\right)  \delta\left(  \mathbf{R}^{\prime
}-j\right)  $ can be divided, regarding which particles appear in a scattering
sequence. To perform this division, from all scattering sequences $\sum
_{i=1}^{N}\sum_{j=1}^{N}\delta\left(  \mathbf{R}-i\right)  \mathbf{T}%
_{ij}\left(  \mathbf{r},\mathbf{r}^{\prime};X\right)  \delta\left(
\mathbf{R}^{\prime}-j\right)  ,$ we extract only the scattering sequences
between the particles from the group $C$:%
\begin{align}
&  \mathbf{T}^{\left(  s\right)  }\left(  \mathbf{R,r};\mathbf{R}^{\prime
},\mathbf{r}^{\prime}||C\right) \nonumber\\
&  =\text{all }s-\text{particle scattering sequences from }\sum_{i=1}^{N}%
\sum_{j=1}^{N}\delta\left(  \mathbf{R}-i\right)  \mathbf{T}_{ij}\left(
\mathbf{r},\mathbf{r}^{\prime};X\right)  \delta\left(  \mathbf{R}^{\prime
}-j\right)  ,\nonumber\\
&  \text{which include all particles from }s-\text{particle group }C.
\label{def Ts}%
\end{align}
The above definition allows to represent the cluster expansion of the
scattering series as follows%
\begin{equation}
\sum_{i=1}^{N}\sum_{j=1}^{N}\delta\left(  \mathbf{R}-i\right)  \mathbf{T}%
_{ij}\left(  \mathbf{r},\mathbf{r}^{\prime};X\right)  \delta\left(
\mathbf{R}^{\prime}-j\right)  =\sum_{s=1}^{N}\sum_{C\subset X,\left\vert
C\right\vert =s}\mathbf{T}^{\left(  s\right)  }\left(  \mathbf{R,r}%
;\mathbf{R}^{\prime},\mathbf{r}^{\prime}||C\right)  .
\label{scattering series cluster exp}%
\end{equation}
In the above expression, $\left\vert C\right\vert $ stands for the number of
particles in the group $C$, whereas $\sum_{C\subset X,\left\vert C\right\vert
=s}$ denotes summation over the $s$-particle groups of particles among
$X=\left\{  1,\ldots,N\right\}  $. Number of such $s-$particle groups is given
by the Newton symbol $\binom{N}{s}$.

Average of the equation (\ref{scattering series cluster exp}) over the
probability distribution function, leads to the cluster expansion for the
average scattering series $\mathbf{T}$ given be the equation (\ref{T micro}),%
\begin{equation}
\mathbf{T}\left(  \mathbf{R,r};\mathbf{R}^{\prime},\mathbf{r}^{\prime}\right)
=\left\langle \sum_{s=1}^{N}\sum_{C\subset X,\left\vert C\right\vert
=s}\mathbf{T}^{\left(  s\right)  }\left(  \mathbf{R,r};\mathbf{R}^{\prime
},\mathbf{r}^{\prime}||C\right)  \right\rangle .
\end{equation}
Since all particles are identical, i.e. the probability distribution $p$ is
symmetric with respect to interchange of the positions $\mathbf{R}_{i}$, all
terms with the same number of particles $s$ in the above expression give the
same contribution. Therefore, we simplify the last expression, by taking one
$s-$particle group $C=\left\{  1,\ldots,s\right\}  $ and multiplying it by the
factor $\binom{N}{s}$. It yields%
\begin{equation}
\mathbf{T}\left(  \mathbf{R,r};\mathbf{R}^{\prime},\mathbf{r}^{\prime}\right)
=\left\langle \sum_{s=1}^{N}\binom{N}{s}\mathbf{T}^{\left(  s\right)  }\left(
\mathbf{R,r};\mathbf{R}^{\prime},\mathbf{r}^{\prime}||1\ldots s\right)
\right\rangle .
\end{equation}
Introducing $s$-particle distribution functions defined by%
\begin{equation}
n\left(  1\ldots s\right)  =\frac{N!}{\left(  N-s\right)  !}\int d^{3}%
R_{s+1}\ldots\int d^{3}R_{N}p\left(  1\ldots N\right)  ,
\end{equation}
we obtain the cluster expansion of the response operator $\mathbf{T}$ in the
following form,%
\begin{align}
\mathbf{T}\left(  \mathbf{R,r};\mathbf{R}^{\prime},\mathbf{r}^{\prime}\right)
&  =\sum_{s=1}^{N}\frac{1}{s!}\mathbf{T}^{\left(  s\right)  }\left(
\mathbf{R,r};\mathbf{R}^{\prime},\mathbf{r}^{\prime}\right)
,\label{cluster exp T intermediate A}\\
\mathbf{T}^{\left(  s\right)  }\left(  \mathbf{R,r};\mathbf{R}^{\prime
},\mathbf{r}^{\prime}\right)   &  =\int d^{3}R_{1}\ldots\int d^{3}%
R_{s}n\left(  1\ldots s\right)  \mathbf{T}^{\left(  s\right)  }\left(
\mathbf{R,r};\mathbf{R}^{\prime},\mathbf{r}^{\prime}||1\ldots s\right)  .
\label{cluster exp T intermediate}%
\end{align}
Its thermodynamic limit is achieved by extending of the summation up to
$N=\infty$ and performing the thermodynamic limit of the $s-$particle
distribution functions $n$. From now on, we will consider the suspension in
the thermodynamic limit.

\subsection{Nodal line}

To perform the cluster expansion of $\mathbf{T}^{\text{irr}}$ operator,
Felderhof, Ford and Cohen used the relation (\ref{Tirr and T}), which may be
represented in the following form%
\begin{equation}
\mathbf{T}^{\text{irr}}=\mathbf{T-TGT+TGTGT-\ldots.} \label{Tirr by T series}%
\end{equation}
Let us look at the second term, i.e. $\mathbf{TGT}$. Representing the
$\mathbf{T}$ by the cluster expansion (\ref{cluster exp T intermediate A}),
produces many terms, each of the form%
\begin{equation}
\mathbf{T}^{\left(  s_{1}\right)  }\mathbf{GT}^{\left(  s_{2}\right)  }.
\end{equation}
In the expression $\mathbf{T}^{\left(  s_{1}\right)  }\mathbf{GT}^{\left(
s_{2}\right)  },$ the scattering sequences between $s_{1}$ particles appearing
in $\mathbf{T}^{\left(  s_{1}\right)  }$ are "connected" by the Green function
$\mathbf{G}$ with the scattering sequences consisted of $s_{2}$ particles
appearing in $\mathbf{T}^{\left(  s_{2}\right)  }$. Altogether, $\mathbf{T}%
^{\left(  s_{1}\right)  }\mathbf{GT}^{\left(  s_{2}\right)  }$ forms
$s_{1}+s_{2}$-particle scattering sequences. The scattering sequences built
from the $\mathbf{T}^{\left(  s_{1}\right)  }\mathbf{GT}^{\left(
s_{2}\right)  }$ are of a special type, i.e. there is a line $\mathbf{G}$
connecting a particle from $\mathbf{T}^{\left(  s_{1}\right)  }$ to a particle
from $\mathbf{T}^{\left(  s_{2}\right)  }$. This line $\mathbf{G}$ is the only
"connection" between the particles from $\mathbf{T}^{\left(  s_{1}\right)  }$
and $\mathbf{T}^{\left(  s_{2}\right)  }$. It is critical to distinguish the
lines $\mathbf{G}$, which are the only connections between some groups of the
particles in a scattering sequence. Those $\mathbf{G}$ are called the nodal
lines \cite{diag_ostateczna_wersja}. It is described by the following
examples. In the scattering sequence given by expression (\ref{scat seq ex1}),
there is one propagator $\mathbf{G}$. It is a nodal line, because it is the
only connection between the groups of particles $\left\{  1\right\}  $ and
$\left\{  2\right\}  $. In the scattering sequence (\ref{scat seq ex2}), there
are three propagators $\mathbf{G}$. The underlined propagator,%
\begin{equation}
\mathbf{M}\left(  1\right)  \underline{\mathbf{G}}\mathbf{M%
}\left(  2\right)  \mathbf{GM}\left(  3\right)  \mathbf{GM}\left(
2\right)  ,
\end{equation}
is a nodal line, because it is the only connection between the group $\left\{
1\right\}  $ and the group $\left\{  23\right\}  $. In diagrammatic language,
the last scattering sequence reads%
\begin{equation}
\text{%
\parbox[c]{3.6cm}{
\begin{tikzpicture}
\linprzer{0}{0}{1.3} \linprzer{0}{1}{1.3} \linprzer{0}{2}{1.3}
\nodalna{0.35}{0}{1} \nodalna{.65}{1}{1} \nodalna{.95}{1}{1}
\operM{.2}{0}  \operM{.5}{1}  \operM{0.8}{2}  \operM{1.1}{1}
\draw[<-] (0.45,0.55) -- (1.2,0.45);
\draw(1.2,0.45) node[anchor=west] {nodal line};
\draw(0,0) node[anchor=east] {1};
\draw(0,1) node[anchor=east] {2};
\draw(0,2) node[anchor=east] {3};
\end{tikzpicture}
}%
.}%
\end{equation}
It is easy to identify a nodal line in diagrammatic language: if cutting a
line of a propagator $\mathbf{G,}$ divides the diagram into two separate
pieces, then the propagator $\mathbf{G}$ is a nodal line.

\subsection{Cluster expansion of $\mathbf{T}$ with nodal lines specified}

In the previous section, we indicated, that an important element of the
analysis of $\mathbf{T}^{\text{irr}}$ is the notion of the nodal line.
Therefore, we perform further division of the scattering series $\mathbf{T,}$
by specifying the nodal lines in the scattering sequences.

In agreement with the definition (\ref{def Ts}), $\mathbf{T}^{\left(
s\right)  }\left(  \mathbf{R,r};\mathbf{R}^{\prime},\mathbf{r}^{\prime
}||1\ldots s\right)  $ represents infinitely many $s-$particle scattering
sequences. We divide them into disjoint sets, by specifying the number of the
nodal lines and by specifying the groups of particles separated by the nodal
lines in the scattering sequence. Those sets are characterized by the number
of groups $g$, the number of particles in each group $s_{1}=\left\vert
C_{1}\right\vert ,\ldots,s_{g}=\left\vert C_{g}\right\vert $, and by saying
which particles from $\left\{  1,\ldots,s\right\}  $ are in the group $C_{i}$.
The number of the groups $g$ is larger by one than the number of the nodal
lines. Since we consider the $s$-particle scattering sequences, we have the
condition $s_{1}+\ldots+s_{g}=s$. We extract from all $s-$particle scattering
sequences $\mathbf{T}^{\left(  s\right)  }\left(  \mathbf{R,r};\mathbf{R}%
^{\prime\prime},\mathbf{r}^{\prime\prime}||1\ldots s\right)  ,$ the scattering
sequences with specified groups of particles $C_{1},\ldots,C_{g}$ separated by
nodal lines, defining%
\begin{align}
&  \mathbf{\bar{T}}^{\left(  s\right)  }\left(  \mathbf{R,r};\mathbf{R}%
^{\prime},\mathbf{r}^{\prime}||C_{1}|\ldots|C_{g}\right)
\nonumber\\
&  =\text{all scattering sequences from }\mathbf{T}^{\left(  s\right)
}\left(  \mathbf{R,r};\mathbf{R}^{\prime},\mathbf{r}^{\prime}||C_{1}\ldots
C_{g}\right)  \text{ with }g-1\text{ nodal lines}\nonumber\\
&  \text{separating particles from the groups }C_{1},\ldots,C_{g}.
\label{def block structure}%
\end{align}
With the above definition, the $s-$particle scattering sequences can be
divided as follows,%
\begin{equation}
\mathbf{T}^{\left(  s\right)  }\left(  \mathbf{R,r};\mathbf{R}^{\prime
},\mathbf{r}^{\prime}||C\right)  =\sum_{g=1}^{s}\sum_{s_{1}+\ldots
+s_{g}=s}\sum_{\substack{C_{1},\ldots,C_{g}\subset C,\\\left\vert
C_{1}\right\vert +\ldots+\left\vert C_{g}\right\vert =s}}\mathbf{\bar{T}%
}^{\left(  s\right)  }\left(  \mathbf{R,r};\mathbf{R}^{\prime
},\mathbf{r}^{\prime}||C_{1}|\ldots|C_{g}\right)  .
\label{Ts struktura blokowa}%
\end{equation}
Here, $\sum_{\substack{C_{1},\ldots,C_{g}\subset C,\\\left\vert C_{1}%
\right\vert =s_{1};\ldots;\left\vert C_{g}\right\vert =s_{g}}}$ denotes
summation over all possible divisions of the set of $s$ particles $\left\{
1,\ldots,s\right\}  $ into $g$ subsets, with $s_{1}$ particles in the first
subset, $s_{2}$ particles in the second subset, etc. There are $s!/\left(
s_{1}!\ldots s_{g}!\right)  $ possible divisions.

Let us now consider the lowest order term $\mathbf{\bar{T}}^{\left(  s\right)
}\left(  C\right)  $ in the expression (\ref{Ts struktura blokowa}), i.e. the
term with $g=1,$ which have no nodal lines. The definition
(\ref{def block structure}) implies that%
\begin{equation}
\mathbf{\bar{T}}^{\left(  s\right)  }\left(  \mathbf{R,r};\mathbf{R}^{\prime
},\mathbf{r}^{\prime}||C\right)  =\text{all scattering sequences from
}\mathbf{T}^{\left(  s\right)  }\left(  \mathbf{R,r};\mathbf{R}^{\prime
},\mathbf{r}^{\prime}||C\right)  \text{ without nodal lines.}\nonumber
\end{equation}
The above scattering sequences without nodal lines play significant role.
They are called the irreducible scattering sequences
\cite{diag_ostateczna_wersja}.

The second order term in the expression (\ref{Ts struktura blokowa}) is the
term with $g=2,$%
\begin{equation}
\sum_{s_{1}+s_{2}=s}\sum_{\substack{C_{1},C_{2}\subset C,\\\left\vert
C_{1}\right\vert +\left\vert C_{2}\right\vert =s}}\mathbf{\bar{T}}^{\left(
s\right)  }\left(  \mathbf{R,r};\mathbf{R}^{\prime},\mathbf{r}^{\prime}%
||C_{1}|C_{2}\right)  .
\end{equation}
From the definition (\ref{def block structure}), it follows, that
$\mathbf{\bar{T}}^{\left(  s\right)  }\left(  C_{1}|C_{2}\right)  $ has one
nodal line $\mathbf{G}$ separating the particles from the groups $C_{1}$ and
$C_{2}$. Therefore, all scattering sequences in $\mathbf{\bar{T}}^{\left(
s\right)  }\left(  C_{1}|C_{2}\right)  $ have the following structure: first -
there are some reflections between the particles from the group $C_{1}$, then
- there is exactly one reflection $\mathbf{G}$ to a particle in the group
$C_{2}$ (nodal line), and then - there are reflections between the particles
from the group $C_{2}$. The reflections between the particles from the group
$C_{1}$ must be irreducible (without a nodal line). The same holds for the
group $C_{2}$. It suggests, that $\mathbf{\bar{T}}^{\left(  s\right)  }\left(
C_{1}|C_{2}\right)  $ has the following structure,%
\begin{equation}
\mathbf{\bar{T}}^{\left(  s\right)  }\left(  C_{1}|C_{2}\right)
=\mathbf{\bar{T}}^{\left(  s_{1}\right)  }\left(  C_{1}\right)  \mathbf{G\bar
{T}}^{\left(  s_{2}\right)  }\left(  C_{2}\right)  ,
\end{equation}
in which the irreducible scattering sequences $\mathbf{\bar{T}}^{\left(
s\right)  }\left(  C\right)  $ appear and the nodal line is written
explicitly. The above formula can be simply proved using the definition
(\ref{def block structure}). Similar results hold for the higher terms of the
expansion (\ref{Ts struktura blokowa}),%
\begin{equation}
\mathbf{\bar{T}}^{\left(  s\right)  }\left(  C_{1}|\ldots|C_{g}\right)
=\mathbf{\bar{T}}^{\left(  s_{1}\right)  }\left(  C_{1}\right)
\mathbf{G\ldots G\bar{T}}^{\left(  s_{g}\right)  }\left(  C_{g}\right)  ,
\end{equation}
for the groups $C_{1},\ldots,C_{g},$ including $s_{1},\ldots,s_{g}$ particles
respectively. In the above formula, the nodal lines separate different
irreducible sections $\mathbf{\bar{T}}^{\left(  s_{i}\right)  }\left(
C_{i}\right)  $ of the scattering sequences $\mathbf{\bar{T}}^{\left(
s\right)  }\left(  C_{1}|\ldots|C_{g}\right)  $. Each irreducible section
$\mathbf{\bar{T}}^{\left(  s_{i}\right)  }\left(  C_{i}\right)  $ is referred
to as "the block". Moreover, by "the block structure" - we mean the way
particles are distributed in the blocks. A block structure is specified as
follows: $C_{1}|\ldots|C_{g}$. The block structure of the scattering sequence
given by the expression (\ref{scat seq ex2}) is $1|23$. It is convenient to
introduce the following notation for the irreducible scattering sequences,%
\begin{equation}
\mathbf{S}_{I}\left(  \mathbf{R,r};\mathbf{R}^{\prime},\mathbf{r}^{\prime
}||C\right)  =\frac{1}{s!}\mathbf{\bar{T}}^{\left(  s\right)  }\left(
\mathbf{R,r};\mathbf{R}^{\prime},\mathbf{r}^{\prime}||C\right)  ,
\end{equation}
with the factor $s!$ Using the above two formulas, we rewrite the expansion
(\ref{Ts struktura blokowa}) as follows%
\begin{equation}
\mathbf{T}^{\left(  s\right)  }\left(  C\right)  =\sum_{g=1}^{s}\sum
_{s_{1}+\ldots+s_{g}=s}\sum_{\substack{C_{1},\ldots,C_{g}\subset
C,\\\left\vert C_{1}\right\vert +\ldots+\left\vert C_{g}\right\vert =s}%
}s_{1}!\ldots s_{g}!\mathbf{S}_{I}\left(  C_{1}\right)  \mathbf{G\ldots
GS}_{I}\left(  C_{g}\right)  . \label{sekwencje na strukture nodalna}%
\end{equation}
The above representation used in the cluster expansion of $\mathbf{T}$
operator represented by equations (\ref{cluster exp T intermediate A}) and
(\ref{cluster exp T intermediate}), after a simple algebra, leads to the
expression%
\begin{equation}
\mathbf{T}=\sum_{g=1}^{\infty}\sum_{C_{1},\ldots,C_{g}}\int dC_{1}\ldots
dC_{g}\ n(C_{1},\ldots,C_{g})\mathbf{S}_{I}(C_{1})\mathbf{G}\ldots
\mathbf{GS}_{I}(C_{g}). \label{T mikro}%
\end{equation}
Cancellation of the factors $s_{i}!$ results from the symmetry of the
probability distribution density $p$ and the fact, that the particles are identical.
The symbol $\sum_{C_{1},\ldots,C_{g}}\int dC_{1}\ldots \int dC_{g}$ denotes
summation over different numbers of particles in each of $g$ groups
and integration over the positions of particles as follows
\begin{eqnarray}
	& \displaystyle \sum_{C_{1},\ldots,C_{g}}\int dC_{1}\ldots dC_{g} f(C_1,\ldots,C_g)  = & \nonumber \\
	& \displaystyle \sum_{n_1=1}^{\infty} \ldots \sum_{n_g=1}^{\infty} \int d^3 R^{1}_{1} \ldots d^3 R^{1}_{n_1} \ldots d^3 R^{g}_{1} \ldots d^3 R^{g}_{n_g}
	\ f(\mathbf{R}^{1}_{1},\ldots,\mathbf{R}^{1}_{n_1},\ldots,\mathbf{R}^{g}_{1},\ldots,\mathbf{R}^{g}_{n_g})
	\label{notation}
\end{eqnarray}

\subsection{Cluster expansion of $\mathbf{T}^{\text{irr}}$}

We go back to the expression (\ref{Tirr by T series}),%
\begin{equation}
\mathbf{T}^{\text{irr}}=\mathbf{T-TGT+TGTGT-\ldots,}
\label{Tirr by T series 2}%
\end{equation}
in order to derive the cluster expansion of the $\mathbf{T}^{\text{irr}}$
operator. As we discussed before, the average scattering series $\mathbf{T}$
includes all possible scattering sequences. $\mathbf{TGT}$ in the equation
(\ref{Tirr by T series 2}) produces scattering sequences with at least one
nodal line, $\mathbf{TGTGT}$ with at least two nodal lines, etc. The analysis
of the above equation relies on a consideration of the scattering sequences
with given number of nodal lines. Therefore, we will consider terms with
different block structures $C_{1}|\ldots|C_{g}$ on the right-hand side of the
equation (\ref{Tirr by T series 2}).

Let us focus first on the block structure $C_{1}$, i.e. on the scattering
sequences without nodal lines. They appear only in the first term,
$\mathbf{T,}$ of the equation (\ref{Tirr by T series 2}), because the higher
terms, $\mathbf{TGT}$, $\mathbf{TGTGT,\ldots,}$ include at least one nodal
line. In the expression (\ref{T mikro}) for $\mathbf{T}$, the irreducible
scattering sequences come only from the term $g=1$. Therefore, all terms on
the right hand-side of the equation (\ref{Tirr by T series 2}) with the block
structure $C_{1},$ are of the form%
\begin{equation}
n(C_{1})\mathbf{S}_{I}(C_{1}).
\end{equation}
Next, we consider the terms on the right-hand side of the equation
(\ref{Tirr by T series 2}) with the block structure $C_{1}|C_{2},$ i.e. with
one nodal line. Such scattering sequences, i.e. $\mathbf{S}_{I}(C_{1}%
)\mathbf{GS}_{I}(C_{2})$, appear only in the first term, $\mathbf{T}$, and in
the second term, $\mathbf{TGT}$. A\ contribution from the $\mathbf{T}$ comes
from the term with $g=2$ of the equation (\ref{T mikro}) and is of the form
$n\left(  C_{1}C_{2}\right)  \mathbf{S}_{I}(C_{1})\mathbf{GS}_{I}(C_{2})$. A
contribution from $\mathbf{TGT}$ has a form of $n\left(  C_{1}\right)
n\left(  C_{2}\right)  \mathbf{S}_{I}(C_{1})\mathbf{GS}_{I}(C_{2})$ and is
produced by the terms with $g=1$ in both $\mathbf{T}$. Those two terms
altogether yield%
\begin{equation}
\left(  n\left(  C_{1},C_{2}\right)  -n\left(  C_{1}\right)  n\left(
C_{2}\right)  \right)  \mathbf{S}_{I}(C_{1})\mathbf{GS}_{I}(C_{2}).
\end{equation}
For the block structure consisted of the three groups $C_{1}|C_{2}|C_{3},$
analysis of the right-hand side of the expression (\ref{Tirr by T series 2})
leads to the contribution of the following form%
\begin{equation}
\left(  n\left(  C_{1}C_{2}C_{3}\right)  -n\left(  C_{1}C_{2}\right)  n\left(
C_{3}\right)  -n\left(  C_{1}\right)  n\left(  C_{2}C_{3}\right)  +n\left(
C_{1}\right)  n\left(  C_{2}\right)  n\left(  C_{3}\right)  \right)
\mathbf{S}_{I}(C_{1})\mathbf{GS}_{I}(C_{2})\mathbf{GS}_{I}(C_{3}).
\end{equation}
\qquad

In such manner, analysis of all block structures $C_{1}|\ldots|C_{g}$ is
possible. Functions appearing along with the block structures of the form
$\mathbf{S}_{I}(C_{1})\mathbf{G}\ldots\mathbf{GS}_{I}(C_{g}),$ are denoted by
$b\left(  C_{1}|\ldots|C_{g}\right)  $ and are called the block distribution
functions \cite{felderhof1982cluster}. Therefore we have%
\begin{align}
b\left(  C_{1}\right)   &  =n(C_{1}),\\
b\left(  C_{1}|C_{2}\right)   &  =n\left(  C_{1},C_{2}\right)  -n\left(
C_{1}\right)  n\left(  C_{2}\right)  ,\\
b\left(  C_{1}|C_{2}|C_{3}\right)   &  =n\left(  C_{1}C_{2}C_{3}\right)
-n\left(  C_{1}C_{2}\right)  n\left(  C_{3}\right)  -n\left(  C_{1}\right)
n\left(  C_{2}C_{3}\right)  +n\left(  C_{1}\right)  n\left(  C_{2}\right)
n\left(  C_{3}\right)  ,\label{b3 przez n}%
\end{align}
for the block structures up to three groups. Expressions for the block
distribution functions for higher number of groups are more and more
complicated. Nevertheless, the block distribution functions may be calculated
from the following recursive formula \cite{michels1989convergence,
diag_ostateczna_wersja}
\begin{subequations}
\label{block distribution functions}%
\begin{align}
b\left(  C\right)   &  =n\left(  C\right)  \\
b\left(  C_{1}|\ldots|C_{k}C_{k+1}|\ldots|C_{g}\right)   &  =b\left(
C_{1}|\ldots|C_{k}\right)  b\left(  C_{k+1}|\ldots|C_{g}\right)  +\nonumber\\
&  +b\left(  C_{1}|\ldots|C_{k}|C_{k+1}|\ldots|C_{g}\right)  .
\end{align}

The above analysis leads to the following cluster expansion of $\mathbf{T}%
^{\text{irr}}$ operator:%
\end{subequations}
\begin{equation}
\mathbf{T}^{\text{irr}}=\sum_{g=1}^{\infty}\sum_{C_{1},\ldots,C_{g}}\int
dC_{1}\ldots dC_{g}b(C_{1}|\ldots|C_{g})\mathbf{S}_{I}(C_{1})\mathbf{G}%
\ldots\mathbf{GS}_{I}(C_{g}). \label{irr mikro}%
\end{equation}
It is worth noting, that the structure of $\mathbf{T}^{\text{irr}}$ operator
is similar to the structure of $\mathbf{T}$ operator given by the expression
(\ref{T mikro}). The only difference lies in the distribution functions: in
$\mathbf{T}^{\text{irr}}$ - the block distribution functions $b\left(
C_{1}|\ldots|C_{g}\right)  $ appear, whereas in $\mathbf{T}$ operator - the
standard $s$-particle distribution functions $n\left(  C_{1}\ldots
C_{g}\right)  $ appear.

\subsection{Self scattering sequences}

There are phenomena in the physics of suspensions, in which only a part of the
scattering series $\mathbf{T,}$ given by equation (\ref{T micro}), plays a
role. An example of the above is the self-diffusion coefficient. It is related
only to those scattering sequences in $\mathbf{T,}$ which start and end at the
same particle. The scattering sequences, which start and end at the same
particle, we call the self-scattering sequences. The self-scattering sequences
are irreducible, because there are no nodal lines in any scattering sequence,
which starts and ends at the same particle. Therefore, the self-scattering
sequences $\mathbf{S}_{I}^{\text{self}}$ are related to the irreducible
scattering sequences $\mathbf{S}_{I}$, as follows,%
\begin{align}
\mathbf{S}_{I}^{\text{self}}\left(  \mathbf{R,r};\mathbf{R}^{\prime
},\mathbf{r}^{\prime}||C\right)   &  =\text{all scattering sequences from
}\mathbf{S}_{I}(\mathbf{R,r};\mathbf{R}^{\prime},\mathbf{r}^{\prime
}||C),\nonumber\\
&  \text{which start and end at the same particle.}%
\end{align}
The average of the self-scattering sequences will be denoted by $\mathbf{B,}$%
\begin{equation}
\mathbf{B}\left(  \mathbf{R,r};\mathbf{R}^{\prime},\mathbf{r}^{\prime}\right)
=\sum_{C_{1}}\int dC_{1}\ n(C_{1})\mathbf{S}_{I}^{\text{self}}(\mathbf{R,r}%
;\mathbf{R}^{\prime},\mathbf{r}^{\prime}||C_{1}). \label{def B}%
\end{equation}
With the above quantity, the scattering sequences in $\mathbf{T}$ can be
divided into the self-scattering sequences and the scattering sequences, which
start and end at different particles (off-scattering sequences). The former
are given by $\mathbf{B}\left(  \mathbf{R,r};\mathbf{R}^{\prime}%
,\mathbf{r}^{\prime}\right)  $, the latter are denoted by $\mathbf{T}%
_{\text{off}}$. Therefore,%
\begin{equation}
\mathbf{T}=\mathbf{B}+\mathbf{T}_{\text{off}}. \label{T into self off}%
\end{equation}
Operator $\mathbf{T}^{\text{irr}}$ can be divided in a similar manner,%
\begin{equation}
\mathbf{T}^{\text{irr}}=\mathbf{B}+\mathbf{T}_{\text{off}}^{\text{irr}}.
\label{Tirr into self off}%
\end{equation}
As shown in the above formulas, the self-part of both operators $\mathbf{T}$
and $\mathbf{T}^{\text{irr}}$ are the same.

It is worth noting, that the self-scattering sequences $\mathbf{B}$ can be
calculated from off-scattering sequences $\mathbf{T}_{\text{off}}$ as follows
\cite{wertheim1973dielectric}%
\begin{equation}
\mathbf{B}\left(  \mathbf{R,r;R}^{\prime}\mathbf{,r}^{\prime}\right)
=n_{1}\delta\left( \mathbf{R-R}^{\prime}\right)\mathbf{M\left(  \mathbf{R,r,r}^{\prime}\right)  +}\delta\left(
\mathbf{R-R}^{\prime}\right)  \left[  \mathbf{T}_{\text{off}}\mathbf{GM%
}\right]  \left(  \mathbf{R,r;R,r}^{\prime}\right)  . \label{self by off}%
\end{equation}
The second term in the above expression corresponds to scattering sequences of
the following structure. The sequences start at the particle at position
$\mathbf{R,}$ then go with all possible scattering sequences $\mathbf{T}%
_{\text{off}}$ to some particle, and then they come back with one reflection
$\mathbf{G}$ to the starting particle at $\mathbf{R}$.
Moreover, in the above equation $n_1$ stands for the one-particle distribution function.

\section{Ring expansion of $\mathbf{T}^{\text{irr}}$}

The cluster expansion (\ref{irr mikro}) of the response operator
$\mathbf{T}^{\text{irr}}$ was derived by Felderhof, Ford and Cohen about three
decades ago \cite{felderhof1982cluster}. This achievement allows to express
the transport coefficients of dispersive media, such as suspensions and
dielectrics, by absolutely convergent integrals. Despite this important step
done by the three scientists, no reasonable statistical physics method of
calculation of transport properties of suspensions can be found in current
literature. We have already discussed this point in the introduction.

In order to give a motivation of our approach introduced further in this
article, we invoke the effective Green function $\mathbf{G}_{\text{eff}}$
defined by%

\begin{equation}
\mathbf{G}_{\text{eff}}=\mathbf{G}+\mathbf{GTG.} \label{Geff}%
\end{equation}
In the case of homogeneous system in the thermodynamic limit, the effective
Green function $\mathbf{G}_{\text{eff}}$ of suspension is related to the Green
function $\mathbf{G}$ of pure fluid according to the following formula,
\cite{diag_ostateczna_wersja}%
\begin{equation}
\mathbf{G}_{\text{eff}}\left(  \mathbf{r},\mathbf{r}^{\prime}\right)
\approx\frac{\eta}{\eta_{\text{eff}}\left(  \phi\right)  }\mathbf{G}\left(
\mathbf{r}-\mathbf{r}^{\prime}\right)  ,\text{ \ \ \ \ \ for }\left\vert
\mathbf{r}-\mathbf{r}^{\prime}\right\vert \rightarrow\infty.
\label{propagator efektywny a oseen}%
\end{equation}
The above formula describes the asymptotic decay of the effective propagator
$\mathbf{G}_{\text{eff}}\left(  \mathbf{r},\mathbf{r}^{\prime}\right)  $ for
the large distances between the points $\mathbf{r}$ and $\mathbf{r}^{\prime}$.
The decay is governed by the effective viscosity of suspension $\eta
_{\text{eff}}\left(  \phi\right)  ,$ which depends on the volume fraction
$\phi=4\pi n_1a^{3}/3,$ with $n_1$ standing for the number density of the
particles in suspension.

Let us consider relation (\ref{Tirr and T}) between $\mathbf{T}$ and
$\mathbf{T}^{\text{irr}},$ which can be inverted and then represented in the
following way,%
\begin{equation}
\mathbf{T}=\mathbf{T}^{\text{irr}}+\mathbf{T}^{\text{irr}}\mathbf{G}%
_{\text{eff}}\mathbf{T}^{\text{irr}},
\end{equation}
where the formula for the effective propagator $\mathbf{G}_{\text{eff}%
}=\mathbf{G}\left(  1-\mathbf{T}^{\text{irr}}\mathbf{G}\right)  ^{-1}$ is
used. Considering only the off-scattering sequences in the above expression,
we receive the equation%
\begin{equation}
\mathbf{T}_{\text{off}}\mathbf{=T}_{\text{off}}^{irr}+\mathbf{T}^{\text{irr}%
}\mathbf{G}_{\text{eff}}\mathbf{T}^{\text{irr}},
\end{equation}
obtained with the application of the formulas (\ref{Tirr into self off}) and
(\ref{T into self off}). Let us notice, that $\mathbf{T}_{\text{off}}\left(
\mathbf{R,r};\mathbf{R}^{\prime},\mathbf{r}^{\prime}\right)  =0$ for
overlapping configurations, i.e. for $\left\vert \mathbf{R}-\mathbf{R}%
^{\prime}\right\vert <2a$. It results from the assumption, that the hard
spheres in suspension cannot overlap and in the expression (\ref{T mikro}),
the distribution function $n\left(  C\right)  $ vanishes for the overlapping
configurations. The vanishing of $\mathbf{T}_{\text{off}}\left(
\mathbf{R,r};\mathbf{R}^{\prime},\mathbf{r}^{\prime}\right)  $ for the
overlapping configurations has a consequence in the last equation. It reduces
to%
\begin{equation}
f\left(  \mathbf{R}-\mathbf{R}^{\prime}\right)  \mathbf{T}_{\text{off}}%
^{irr}\left(  \mathbf{R,r};\mathbf{R}^{\prime},\mathbf{r}^{\prime}\right)
=-f\left(  \mathbf{R}-\mathbf{R}^{\prime}\right)  \left[  \mathbf{T}%
^{\text{irr}}\mathbf{G}_{\text{eff}}\mathbf{T}^{\text{irr}}\right]  \left(
\mathbf{R,r};\mathbf{R}^{\prime},\mathbf{r}^{\prime}\right)  ,
\label{ring overlap}%
\end{equation}
after its multiplication by a function $f\left(  \mathbf{R}-\mathbf{R}%
^{\prime}\right)  $ defined by%
\begin{equation}
f\left(  \mathbf{R}-\mathbf{R}^{\prime}\right)  =\left\{
\begin{array}
[c]{cc}%
1 & \text{for\ }\left\vert \mathbf{R}-\mathbf{R}^{\prime}\right\vert <2a\\
0 & \text{for\ }\left\vert \mathbf{R}-\mathbf{R}^{\prime}\right\vert \geq2a
\end{array}
\right.  , \label{mayer function}%
\end{equation}
which equals $0$ for nonoverlapping and $1$ for overlapping configurations of
the two particles centered at $\mathbf{R}$ and $\mathbf{R}^{\prime}$.

The equation (\ref{ring overlap}) shows, that in $\mathbf{T}^{\text{irr}},$
there are some contributions with the effective propagator $\mathbf{G}%
_{\text{eff}}$. It suggests, that the propagator $\mathbf{G}$ appearing in the
cluster expansion (\ref{irr mikro}) of Felderhof, Ford and Cohen can be
renormalized. In other words, $\mathbf{T}^{\text{irr}}$ can be given by the
following formula,%
\begin{equation}
\mathbf{T}^{\text{irr}}=\sum_{d=1}^{\infty}\sum_{C_{1}\ldots C_{d}}\int
dC_{1}\ldots dC_{d}H(C_{1}|\ldots|C_{d})\mathbf{S}_{I}(C_{1})\mathbf{G}%
_{\text{eff}}\ldots\mathbf{G}_{\text{eff}}\mathbf{S}_{I}(C_{d}),
\label{Tirr rozw pierscieniowe}%
\end{equation}
with yet unknown functions $H\left(  C_{1}|\ldots|C_{d}\right)  ,$ which we
call the block correlation functions. The above expression for $\mathbf{T}%
^{\text{irr}}$ has the same structure as expression (\ref{irr mikro}), but
contains the effective propagator $\mathbf{G}_{\text{eff}},$ instead of the
propagator $\mathbf{G}$ and also contains the block correlation functions
$H\left(  C_{1}|\ldots|C_{d}\right)  ,$ instead of the block distribution
functions $b\left(  C_{1}|\ldots|C_{g}\right)  $. We call the expression
(\ref{Tirr rozw pierscieniowe}) for $\mathbf{T}^{\text{irr}}$ operator, the
ring expansion, in order to differentiate it from the cluster expansion
(\ref{irr mikro}) of this operator introduced by Felderhof, Ford and Cohen.

Below, we prove the ring expansion (\ref{Tirr rozw pierscieniowe}) and derive
a formula for the block correlation functions $H\left(  C_{1}|\ldots
|C_{d}\right)  $. We will use the similar approach, as in the derivation of
the Felderhof, Ford and Cohen's formula (\ref{irr mikro}) in the previous
section. We will consider the right-hand sides of both expressions
(\ref{irr mikro}) and (\ref{Tirr rozw pierscieniowe}), considering scattering
sequences with a given block structure $C_{1}|\ldots|C_{g}$.

We start with the block structure $\mathbf{S}_{I}(C_{1})$, i.e. the block
structure without a nodal line. In the expression
(\ref{Tirr rozw pierscieniowe}), the block structure $\mathbf{S}_{I}(C_{1})$
without a nodal line appears only in the term $d=1.$ In the expression
(\ref{irr mikro}), the block structure $\mathbf{S}_{I}(C_{1})$ also appears in
the lowest order term $g=1$ only. Therefore, equality of the expressions
(\ref{irr mikro}) and (\ref{Tirr rozw pierscieniowe}), on the level of the
irreducible scattering sequences $\mathbf{S}_{I}(C_{1}),$ is possible if we
assume%
\begin{equation}
b(C_{1})=H(C_{1}). \label{block correlations H g1}%
\end{equation}

Before further considerations for a general block structure, let us find all
terms in the ring expansion (\ref{Tirr rozw pierscieniowe}), which have the
block structure $C_{1}|C_{2}|C_{3}|C_{4}$. For the purpose of the above, we
need the cluster expansion of the effective Green function $\mathbf{G}%
_{\text{eff}},$%
\begin{equation}
\mathbf{G}_{\text{eff}}=\sum_{g=0}^{\infty}\sum_{C_{1},\ldots,C_{g}}\int
dC_{1}\ldots dC_{g}\ n(C_{1},\ldots,C_{g})\mathbf{GS}_{I}(C_{1})\mathbf{G}%
\ldots\mathbf{GS}_{I}(C_{g})\mathbf{G,} \label{Geff cluster expansion}%
\end{equation}
which is a straightforward consequence of the expressions (\ref{Geff}) and
(\ref{T mikro}). In the above formula, the term with $g=0$ corresponds to the
Oseen tensor $\mathbf{G}$. The effective Green function introduces one, two,
and more nodal lines in the block structure $\mathbf{S}_{I}\left(
C_{i}\right)  \mathbf{G}_{eff}\mathbf{S}_{I}\left(  C_{j}\right)  $.
Therefore, the scattering sequences $\mathbf{S}_{I}\left(  C_{1}\right)
\mathbf{GS}_{I}\left(  C_{2}\right)  \mathbf{GS}_{I}\left(  C_{3}\right)
\mathbf{GS}_{I}\left(  C_{4}\right)  $ appear in the ring expansion
(\ref{Tirr rozw pierscieniowe}) in the term corresponding to $d=2$,%
\begin{equation}
H(C_{i_{1}}|C_{i_{2}})\mathbf{S}_{I}(C_{i_{1}})\mathbf{G}_{\text{eff}%
}\mathbf{S}_{I}(C_{i_{2}}), \label{re d2}%
\end{equation}
in the term $d=3,$%
\begin{equation}
H(C_{i_{1}}|C_{i_{2}}|C_{i_{3}})\mathbf{S}_{I}(C_{i_{1}})\mathbf{G}%
_{\text{eff}}\mathbf{S}_{I}(C_{i_{2}})\mathbf{G}_{\text{eff}}\mathbf{S}%
_{I}(C_{i_{3}}), \label{re d3}%
\end{equation}
and in the term $d=4,$%
\begin{equation}
H(C_{i_{1}}|C_{i_{2}}|C_{i_{3}}|C_{i_{4}})\mathbf{S}_{I}(C_{i_{1}}%
)\mathbf{G}_{\text{eff}}\mathbf{S}_{I}(C_{i_{2}})\mathbf{G}_{\text{eff}%
}\mathbf{S}_{I}(C_{i_{3}})\mathbf{G}_{\text{eff}}\mathbf{S}_{I}(C_{i_{4}}).
\label{re d4}%
\end{equation}
Other terms in the ring expansion (\ref{Tirr rozw pierscieniowe}) do not
contain the scattering sequences with three nodal lines, because the term
$d=1$ contains no nodal lines, whereas the terms corresponding to $d\geq5$
contain at least four nodal lines.

Each of the terms in the expressions (\ref{re d2}-\ref{re d4}) contains many
different scattering sequences produced by the cluster expansion
(\ref{Geff cluster expansion}) of the effective Green function. In the case of
the expression (\ref{re d2}), only the term $n\left(  C_{2}C_{3}\right)
\mathbf{GS}_{I}\left(  C_{2}\right)  \mathbf{GS}_{I}\left(  C_{3}\right)
\mathbf{G}$ in the expansion (\ref{Geff cluster expansion}) produces the block
structure $C_{1}|C_{2}|C_{3}|C_{4},$ yielding%
\begin{equation}
H\left(  C_{1}|C_{4}\right)  n\left(  C_{2}C_{3}\right)  \mathbf{S}_{I}\left(
C_{1}\right)  \mathbf{GS}_{I}\left(  C_{2}\right)  \mathbf{GS}_{I}\left(
C_{3}\right)  \mathbf{GS}_{I}\left(  C_{4}\right)  . \label{pom 1}%
\end{equation}
In order to obtain this, we assume $i_{1}=1$, $i_{2}=4$ in the expression
(\ref{re d2}). In the case of the expression (\ref{re d3}), there are two
possibilities leading to the block structure $C_{1}|C_{2}|C_{3}|C_{4}$. The
first possibility corresponds to the situation, when the first (left)
propagator $\mathbf{G}_{\text{eff}}$ in the expression (\ref{re d3})
introduces one nodal line and the second propagator introduces two nodal
lines. Therefore, we assume $i_{1}=1$, $i_{2}=2$, $i_{3}=4$ and the
contribution of the term given by the expression (\ref{re d3}) is%
\begin{equation}
H\left(  C_{1}|C_{2}|C_{4}\right)  n\left(  C_{3}\right)  \mathbf{S}%
_{I}\left(  C_{1}\right)  \mathbf{GS}_{I}\left(  C_{2}\right)  \mathbf{GS}%
_{I}\left(  C_{3}\right)  \mathbf{GS}_{I}\left(  C_{4}\right)  .
\end{equation}
The second possibility corresponds to an opposite situation, when the first
propagator in expression (\ref{re d3}) introduces two nodal lines and the
second propagator introduces one nodal line. Here, we assume $i_{1}=1$,
$i_{2}=3$, $i_{3}=4$ and the contribution is%
\begin{equation}
H\left(  C_{1}|C_{3}|C_{4}\right)  n\left(  C_{2}\right)  \mathbf{S}%
_{I}\left(  C_{1}\right)  \mathbf{GS}_{I}\left(  C_{2}\right)  \mathbf{GS}%
_{I}\left(  C_{3}\right)  \mathbf{GS}_{I}\left(  C_{4}\right)  .
\end{equation}
In the case of expression (\ref{re d4}), there is only one possibility to
obtain the block structure $C_{1}|C_{2}|C_{3}|C_{4},$ i.e. when all
propagators $\mathbf{G}_{\text{eff}}$ introduce only one nodal line
$\mathbf{G}$. In this case, we have $i_{1}=1,$ $i_{2}=2$, $i_{3}=3$, $i_{4}=4$
and obtain the following contribution,%
\begin{equation}
H\left(  C_{1}|C_{2}|C_{3}|C_{4}\right)  \mathbf{S}_{I}\left(  C_{1}\right)
\mathbf{GS}_{I}\left(  C_{2}\right)  \mathbf{GS}_{I}\left(  C_{3}\right)
\mathbf{GS}_{I}\left(  C_{4}\right)  . \label{pom 4}%
\end{equation}
Finally, all terms in the expression (\ref{Tirr rozw pierscieniowe}), which
have the block structure $C_{1}|C_{2}|C_{3}|C_{4}$ (i.e. terms given by
expressions (\ref{pom 1}-\ref{pom 4})), after comparison with the term
containing the same block structure $C_{1}|C_{2}|C_{3}|C_{4}$ from the
equation (\ref{irr mikro}), lead to equality%
\begin{align}
b\left(  C_{1}|C_{2}|C_{3}|C_{4}\right)   &  =H\left(  C_{1}|C_{4}\right)
n\left(  C_{2}C_{3}\right)  +H\left(  C_{1}|C_{2}|C_{4}\right)  n\left(
C_{3}\right) \nonumber\\
&  +H\left(  C_{1}|C_{3}|C_{4}\right)  n\left(  C_{2}\right)  +H\left(
C_{1}|C_{2}|C_{3}|C_{4}\right)  . \label{block correlations H g4}%
\end{align}

Similar considerations for the block structures $C_{1}|C_{2}$ and $C_{1}%
|C_{2}|C_{3}$ lead to the expressions%
\begin{equation}
b\left(  C_{1}|C_{2}\right)  =H\left(  C_{1}|C_{2}\right)  ,
\label{rownosc h2 b2}%
\end{equation}%
\begin{equation}
b\left(  C_{1}|C_{2}|C_{3}\right)  =H\left(  C_{1}|C_{2}|C_{3}\right)
+H\left(  C_{1}|C_{3}\right)  n\left(  C_{2}\right)  . \label{b3 przez H}%
\end{equation}

The above considerations for the block structure $C_{1}|C_{2}|C_{3}|C_{4}$,
leading to the formula (\ref{block correlations H g4}), can be generalized to
the case of a block structure $C_{1}|\ldots|C_{g}$ consisted of $g\geq2$
groups. The block structures $C_{1}|\ldots|C_{g}$ appear in the ring
expansion (\ref{Tirr rozw pierscieniowe}) in the terms $d=2,\ldots,g$ only,
i.e. the terms of the form $\mathbf{S}_{I}\left(  C_{1}\right)  \mathbf{G}%
_{eff}\ldots\mathbf{G}_{eff}\mathbf{S}_{I}\left(  C_{d}\right)  $. As the term
(\ref{re d3}) for the case $g=4$ introduces the block structure $C_{1}%
|C_{2}|C_{3}|C_{4}$ in two ways, each of $\mathbf{S}_{I}\left(  C_{1}\right)
\mathbf{G}_{eff}\ldots\mathbf{G}_{eff}\mathbf{S}_{I}\left(  C_{d}\right)  $
can introduce the block structure $C_{1}|\ldots|C_{g}$ in several ways. All
terms can be uniquely classified by specification of $d$ groups $C_{i_{1}%
}|\ldots|C_{i_{d}}$ among $C_{1}|\dots|C_{g},$ which come from the blocks
$\mathbf{S}_{I}$ in the expression $\mathbf{S}_{I}(C_{1})\mathbf{G}%
_{eff}\ldots\mathbf{G}_{eff}\mathbf{S}_{I}(C_{d})$. The edge groups must be
the same, therefore $i_{1}=1$ and $i_{d}=g$. Each set of numbers
$1=i_{1}<i_{2}<\ldots<i_{d-1}<i_{d}=g$ corresponds to a single term in the
expression $\mathbf{S}_{I}(C_{1})\mathbf{G}_{eff}\ldots\mathbf{G}%
_{eff}\mathbf{S}_{I}(C_{d}),$ which produces the block structure
$\mathbf{S}_{I}(C_{1})\mathbf{G}\ldots\mathbf{GS}_{I}(C_{g})$. Comparison of
all terms in the expansion (\ref{Tirr rozw pierscieniowe}) and
(\ref{irr mikro}), producing the block structure $\mathbf{S}_{I}%
(C_{1})\mathbf{G}\ldots\mathbf{GS}_{I}(C_{g})$ yields%
\begin{gather}
b\left(  C_{1}|\ldots|C_{g}\right)  =\nonumber\\
\sum_{d=2}^{g}\sum_{1=i_{1}<i_{2}<\ldots<i_{d}=g}~H(C_{i_{1}}|\ldots|C_{i_{d}%
})\times\nonumber\\
n(\left\{  C_{i_{1}}\ldots C_{i_{2}}\right\}  \backslash\left\{  C_{i_{1}%
}C_{i_{2}}\right\}  )\ldots n\left(  \left\{  C_{i_{d-1}}\ldots C_{i_{d}%
}\right\}  \backslash\left\{  C_{i_{d-1}}C_{i_{d}}\right\}  \right)  ,
\label{bReprPierscieniowa}%
\end{gather}
which is valid for $g\geq2$. The symbol '$\backslash$' denotes a difference of
sets of the particles, e.g. $\left\{  12567\right\}  \backslash\left\{
56\right\}  =\left\{  127\right\}  $. We assume that for the empty set
$n\left(  \varnothing\right)  =1$. The above formula is a recursive expression
for the block correlation functions $H$.

\subsection{Comparison of ring and cluster expansion}

The ring expansion (\ref{Tirr rozw pierscieniowe}) introduced in the previous
section is a rigorous expression for the response operator $\mathbf{T}%
^{\text{irr}}$. This is an alternative formula to the cluster expansion
(\ref{irr mikro}) of Felderhof, Ford and Cohen. The cluster expansion and the
ring expansion have the same structure: the irreducible scattering sequences
$\mathbf{S}_{I}$ connected by the propagators ($\mathbf{G}$ or $\mathbf{G}%
_{\text{eff}}$), are averaged over configurations of particles, weighted with
the distribution functions ($b$ or $H$). From that perspective, and due to
fact, that the effective Green function $\mathbf{G}_{\text{eff}}$ appears in
the ring expansion instead of $\mathbf{G}$, the formula
(\ref{Tirr rozw pierscieniowe}) can be seen as the renormalized cluster expansion.

There are two important differences between the ring and the cluster
expansion. The first difference lies in the propagators. The propagator
$\mathbf{G,}$ which appears in the cluster expansion (\ref{irr mikro}),
includes only information concerning liquid. On the other hand, the effective
Green function $\mathbf{G}_{\text{eff}}$ in the expression
(\ref{Tirr rozw pierscieniowe}), contains macroscopic information about
suspension. It is exhibited by the appearance of the effective viscosity
$\eta_{\text{eff}}$ in the asymptotic form of the effective propagator for
large distances showed in the formula (\ref{propagator efektywny a oseen}).
The effective viscosity $\eta_{\text{eff}}$ of hard-sphere suspension may
significantly differ from the viscosity $\eta$ of pure liquid - especially for
the higher volume fractions $\phi$.

The second difference between the ring and the cluster expansion lies in the
distribution functions. The block distribution functions $b$ appearing in the
cluster expansion are given with the formula
(\ref{block distribution functions}). The block correlation functions $H,$
which appeared in the ring expansion, are defined with the expression
(\ref{bReprPierscieniowa}). There is an essential difference between $b$ and
$H$. It is related to the cluster property of the distribution functions $n,$
which we assume in this article. The cluster property relies on the
factorization of the distribution function $n\left(  C_{1}C_{2}\right)  $ in
the limit of large distance between the groups $C_{1}$ and $C_{2}$,%
\begin{equation}
n\left(  C_{1}C_{2}\right)  \rightarrow n\left(  C_{1}\right)  n\left(
C_{2}\right)  .
\end{equation}
A use of the above cluster property of the distribution functions $n$ in the
equation (\ref{b3 przez n}), when the group $C_{2}$ in the middle of the block
structure $C_{1}|C_{2}|C_{3}$ goes away from the other groups, leads to the
following factorization of the block distribution function%
\begin{equation}
b\left(  C_{1}|C_{2}|C_{3}\right)  \longrightarrow b\left(  C_{1}%
|C_{3}\right)  b\left(  C_{2}\right)  .
\label{pom blokowa fun roz wl grupowa 3}%
\end{equation}
The cluster property of the distribution functions $n$ applied in the equation
(\ref{b3 przez H}) for $H\left(  C_{1}|C_{2}|C_{3}\right)  $ in the same limit
- when the group $C_{2}$ goes away - results in the following decay of the
block correlation function,%
\begin{equation}
H\left(  C_{1}|C_{2}|C_{3}\right)  \longrightarrow0. \label{asymptotyka H3}%
\end{equation}
The above asymptotic decay is a motivation for the name of the block
correlation functions $H$. The above property is also a motivation for the
name of the expression (\ref{Tirr rozw pierscieniowe}) - i.e. ring expansion.
Two subsequent blocks, $\mathbf{S}_{I}\left(  C_{i}\right)  $ and
$\mathbf{S}_{I}\left(  C_{i+1}\right)  ,$ in the ring expansion
(\ref{Tirr rozw pierscieniowe}) are 'connected' by, both, the effective
propagator $\mathbf{G}_{\text{eff}}$ and by the correlation function $H\left(
\ldots|C_{i}|C_{i+1}|\ldots\right)  $ - both 'connections' vanish, when
$C_{i}$ goes away from $C_{i+1}$. We imagine, that such double-connection of
the $\mathbf{S}_{I}\left(  C_{i}\right)  $ and $\mathbf{S}_{I}\left(
C_{i+1}\right)  $ form a 'ring'.

The ring expansion of $\mathbf{T}^{\text{irr}}$
represented by the equation (\ref{Tirr rozw pierscieniowe}) along with the expression (\ref{bReprPierscieniowa})
for the block correlation functions is the main analitycal result of this article.
It is an alternative expression to the Felderhof, Ford and Cohen's cluster expansion represented by the formula (\ref{irr mikro}). 
Our ring expansion appears as a result of a resummation performed on the level of the cluster expansion.
This resummation procedure leads to the ring expansion which has similar structure as the structure of the cluster expansion.
The role of the Oseen tensors in the cluster expansion - after resummation - is played be the effective Green function $\mathbf{G}_{\text{eff}}$
given by the formula (\ref{Geff}).
The effective Green function has a physical interpretation because it relates the force (generating the ambient flow)
with the velocity field of the suspension - in contrast to the Oseen tensor which relates the force with the velocity field of a pure liquid.
Moreover, the effective Green function is related to the effective visicosity, as the expression (\ref{propagator efektywny a oseen}) shows.
Because the resummation procedure leads to a similar structure as the structure of the starting expression,
we call this procedure the renormalization.
Consequently, the effective Green function may also be called the renormalized (effective) Green function.  
The ring expansion is further used in the next section to introduce a new method of calculations of transport properties of suspensions.

The above derivation of the ring expansion is presented in the short-hand notation which emphasizes the idea of the underlying physics.
It is worth presenting the ring expansion without the short-hand notation.
Following the expression (\ref{notation}) the lowest two terms of the ring expansion are given by
\begin{eqnarray}
	& \mathbf{T}^{\text{irr}}\left(  \mathbf{R,r};\mathbf{R}^{\prime},\mathbf{r}^{\prime}\right)=& \nonumber \\
 &\displaystyle \sum_{n_1=1}^{\infty} \int d^3 R^{1}_{1} \ldots d^3 R^{1}_{n_1} 
	H\left( \mathbf{R}^{1}_{1},\ldots,\mathbf{R}^{1}_{n_1} \right) 
	\mathbf{S}_{I}(\mathbf{R,r};\mathbf{R}^{\prime},\mathbf{r}^{\prime} || \mathbf{R}^{1}_{1},\ldots,\mathbf{R}^{1}_{n_1}) \nonumber \\
	& \displaystyle + \sum_{n_1=1}^{\infty}\sum_{n_2=1}^{\infty}
	\int d^3 R^{1}_{1} \ldots d^3 R^{1}_{n_1} d^3 R^{2}_{1} \ldots d^3 R^{2}_{n_2} 
	d^3 r^{\prime \prime } d^3 r^{\prime \prime \prime } d^3 R^{\prime \prime } d^3 R^{\prime \prime \prime }
	H\left( \mathbf{R}^{1}_{1},\ldots,\mathbf{R}^{1}_{n_1}|\mathbf{R}^{2}_{1},\ldots,\mathbf{R}^{2}_{n_2} \right)
	\nonumber \\ & \times 
	\mathbf{S}_{I}(\mathbf{R,r};\mathbf{R}^{\prime \prime},\mathbf{r}^{\prime \prime} || \mathbf{R}^{1}_{1},\ldots,\mathbf{R}^{1}_{n_1})
	\mathbf{G}_{\text{eff}}\left(\mathbf{r}^{\prime \prime},\mathbf{r}^{\prime \prime \prime } \right)
	\mathbf{S}_{I}(\mathbf{R}^{\prime \prime \prime},\mathbf{r}^{\prime \prime \prime } ;\mathbf{R}^{\prime},\mathbf{r}^{\prime} || \mathbf{R}^{2}_{1},\ldots,\mathbf{R}^{2}_{n_2})
	\nonumber \\ &
	+ \ldots
\end{eqnarray}

\section{Renormalization of Clausius-Mossotti approximation}

After the derivation of the cluster expansion (\ref{irr mikro}) of the
$\mathbf{T}^{\text{irr}}$ operator, Felderhof, Ford and Cohen gave the
microscopic explanation of the Clausius-Mossotti formula
\cite{felderhof1983clausius}. It is an expression for the relative dielectric
constant of a nonpolar dielectric system. It may be derived using the
macroscopic considerations \cite{feynman2013feynman}. Felderhof, Ford and
Cohen explained the Clausius-Mossotti formula on the microscopic level,
showing a class of terms in the cluster expansion of $\mathbf{T}^{\text{irr}%
},$ which leads to the Clausius-Mossotti relation.

Going along the line of the explanations of Felderhof, Ford, and Cohen
\cite{felderhof1983clausius} for a dielectric system, the Clausius-Mossotti
relation is obtained, when the operator $\mathbf{T}_{\text{CM}}^{\text{irr}}$
defined with the following formula%
\begin{equation}
\mathbf{T}_{\text{CM}}^{\text{irr}}=\mathbf{T}^{\text{irr}}\left(  1+\left[
h\mathbf{G}\right]  \mathbf{T}^{\text{irr}}\right)  ^{-1}%
,\label{def operator CM}%
\end{equation}
is approximated by the single particle term,%
\begin{equation}
\mathbf{T}_{\text{CM}}^{\text{irr}}\left(  \mathbf{R,r;R}^{\prime}\mathbf{,r}^{\prime}\right)
\approx n_1 \delta\left( \mathbf{R-R}^{\prime}\right) 
\mathbf{M\left(  \mathbf{R,r,r}^{\prime}\right)} .\label{przbl na Tirr_c w CM}%
\end{equation}
In the above definition (\ref{def operator CM}), a superposition between the
quantities $\mathbf{T}^{\text{irr}}\left(  \mathbf{R,r};\mathbf{R}^{\prime
},\mathbf{r}^{\prime}\right)  $ and $\left[  h\mathbf{G}\right]  \left(
\mathbf{R,r;R}^{\prime}\mathbf{,r}^{\prime}\right)  =h\left(  \mathbf{R}%
,\mathbf{R}^{\prime}\right)  \mathbf{G}\left(  \mathbf{r},\mathbf{r}^{\prime
}\right)  $ appears. $h\left(  \mathbf{R},\mathbf{R}^{\prime}\right)  $ stands
for the two-particle correlation function. It is worth noting, that in the
reference \cite{felderhof1983clausius}, instead of the two-particle
correlation function $h\left(  \mathbf{R},\mathbf{R}^{\prime}\right)  $,
function $-f\left(  \mathbf{R},\mathbf{R}^{\prime}\right)  $ with $f$ defined
by the formula (\ref{mayer function}) appears. Both possibilities lead to the
Clausius-Mossotti relation. We call the operator $\mathbf{T}_{\text{CM}%
}^{\text{irr}}$ - the Clausius-Mossotti operator - because it is a
straightforward generalization of the Clausius-Mossotti function. In case
of suspensions, the above procedure leads to the Saito formula for the
effective viscosity \cite{bedeaux1987effective}, namely%
\begin{equation}
\frac{\eta_{\text{eff}}}{\eta}=\frac{1+\frac{3}{2}\phi}{1-\phi}.\label{saito}%
\end{equation}

The Clausius-Mossotti approximation is expressed by the approximated formula
(\ref{przbl na Tirr_c w CM}), applied to the Clausius-Mossotti operator,
defined by the equation (\ref{def operator CM}). In this equation, the Oseen
tensor $\mathbf{G}$ appears. In the previous section, we show that the
propagators $\mathbf{G}$ in the cluster expansion (\ref{irr mikro}) can be
renormalized and, as a result, $\mathbf{T}^{\text{irr}}$ operator can be
represented by the ring expansion (\ref{Tirr rozw pierscieniowe}), with the
effective Green function $\mathbf{G}_{\text{eff}}$ appearing instead of
$\mathbf{G}$. This suggests to define renormalized Clausius-Mossotti operator
$\mathbf{T}_{\text{RCM}}^{\text{irr}},$ as follows,%
\begin{equation}
\mathbf{T}_{\text{RCM}}^{\text{irr}}=\mathbf{T}^{\text{irr}}\left(  1+\left[
h\mathbf{G}_{\text{eff}}\right]  \mathbf{T}^{\text{irr}}\right)  ^{-1},
\label{def ren operator CM}%
\end{equation}
in analogy to the formula (\ref{def operator CM}), and to generalize the
Clausius-Mossotti approximation by%
\begin{equation}
\mathbf{T}_{\text{RCM}}^{\text{irr}}\approx\mathbf{B,} \label{uog przybl CM}%
\end{equation}
in analogy to the approximation (\ref{przbl na Tirr_c w CM}). In the latter
equation, instead of the single particle response operator $\mathbf{M}$
appearing in the Clausius-Mossotti approximation, we take into account all
self-scattering sequences $\mathbf{B,}$ introduced before with the formula
(\ref{def B}). The equations (\ref{def ren operator CM}), (\ref{Tirr and T}),
(\ref{Geff}), (\ref{T into self off}), (\ref{self by off}) along with the
approximation (\ref{uog przybl CM}) define the renormalized Clausius-Mossotti
approximation. Those equations form a close system of equations for operators
$\mathbf{B}$, $\mathbf{T}_{\text{RCM}}^{\text{irr}}$, $\mathbf{T}^{\text{irr}%
}$, $\mathbf{G}_{\text{eff}}$, $\mathbf{T}$, and $\mathbf{T}_{\text{off}}$.
The system can be solved for given volume fraction $\phi$ (or the single
particle density $n_{1}$) and for given two-particle correlation function
$h\left(  \mathbf{R},\mathbf{R}^{\prime}\right)  $. We solve those equations
numerically. Not to interrupt our line of reasoning, we refer the reader to
the appendix \ref{app solution} containing the technical details of our
numerical calculations. From $\mathbf{T}^{\text{irr}}$ found within the
renormalized Clausius-Mossotti approximation, one can calculate further the
short-time transport characteristics, such as the effective viscosity
$\eta_{\text{eff}}$ from the equations (\ref{eta micro 1}-\ref{eta micro 3})
and the wave dependent hydrodynamic function $H\left(  q\right)  $ with the
collective diffusion $D_{c}$ and the self-diffusion $D_{s}$ coefficient from
the equations (\ref{hyd fun mikro}-\ref{self dif}).

Before presenting in the next section the results for the transport characteristics
calculated within the generalized Clausius-Mossotti approximation,
in what follows we comment on its physical meaning.
To this end we discuss the $\mathbf{T}$ operator (see equation (\ref{def T})) obtained within the Clausius-Mossotti approximation denoted by $\mathbf{T}^{CM}$.
The Clausius-Mossotti approximation (\ref{przbl na Tirr_c w CM}) - by equations (\ref{def operator CM}) and (\ref{Tirr and T}) -
leads to the expression
\begin{equation}
	\mathbf{T}^{CM}=n_1\mathbf{M}\left[ 1- [g\mathbf{G}]n_1 \mathbf{M} \right]^{-1},
	\label{T in CM}
\end{equation}
where $g$ is the radial distribution function related to the correlation function $h$ as follows, $g=1+h$.
The above formula for $\mathbf{T}$ can be interpreted in terms of scattering sequences.
The $\mathbf{T}$ operator - which on rigorous level is given by the sum of all possible scattering sequences 
as the equation (\ref{scattering series}) shows - in the Clausius-Mossotti approximation
is given by a sum of scattering sequences in which the reflections never go back to a
particle which already reflected the flow. Moreover, there are correlations $g$ only between neighboring particles
in the scattering sequences.
On the other hand the generalized Clausius-Mossotti approximation (\ref{uog przybl CM})
- along with the rigorous equations (\ref{def ren operator CM}), (\ref{Geff}) and (\ref{Tirr and T}) -
leads to the following formula for the $\mathbf{T}$ operator (denoted by $\mathbf{T}^{RCM}$),
\begin{equation}
	\mathbf{T}^{RCM}=\mathbf{B}\left[ 1- \left( [g\mathbf{G}] + h[GTG] \right) \mathbf{B} \right]^{-1}.
\end{equation}
The above equation differs from the expression (\ref{T in CM}), in particular by the term $h[GTG]$.
This term contains the dominant terms in the virial expansion on three-particle level for the sedimemntation coefficient
(cf. $b_4$ coefficient in the reference \cite{Cichocki2002three})
and for the effective viscosity (cf. $\nu_2$ coefficient in the reference \cite{trojczastkowalepkosc}).
Therefore we expect that the renormalized (generalized) Clausius-Mossotti approximation
will give more accurate results than the original Clausius-Mossotti approximation.


\section{Results and discussion}

\begin{figure}[ptb]%
\includegraphics[
height=2.3592in,
width=3.3676in
]%
{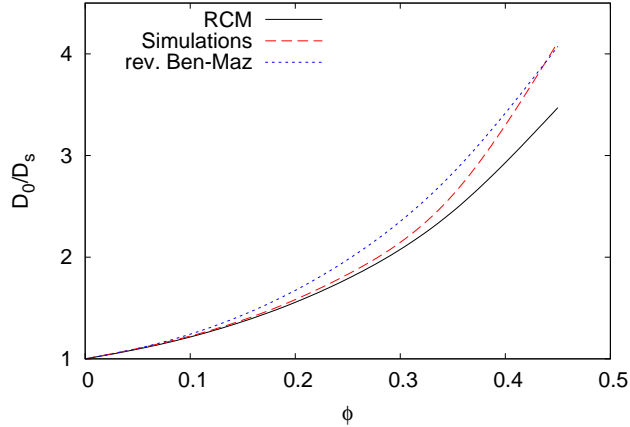}%
\caption{Inverse of the translational short-time self-diffusion coefficient $D_{s}$, equation (\ref{self dif}),
normalized by the self-diffusion coefficient of a single particle, $D_{0}=k_{B}T/\left(  6\pi\eta a\right)  $,
as a function of volume fraction $\phi$
for monodisperse suspension of hard-spheres in equilibrium.
Black (solid line) - the renormalized Clausius-Mossotti approximation introduced in this article,
red (long-dashed) line - numerical simulations of Abade at al. \cite{abade2010short},
blue (short-dashed) line - the revised Beenakker-Mazur method \cite{makuch2012transport}.}%
\label{rys self-diffusion}%
\end{figure}

\begin{figure}[ptbh]%
\includegraphics[
height=5.4982cm,
width=8.4987cm
]%
{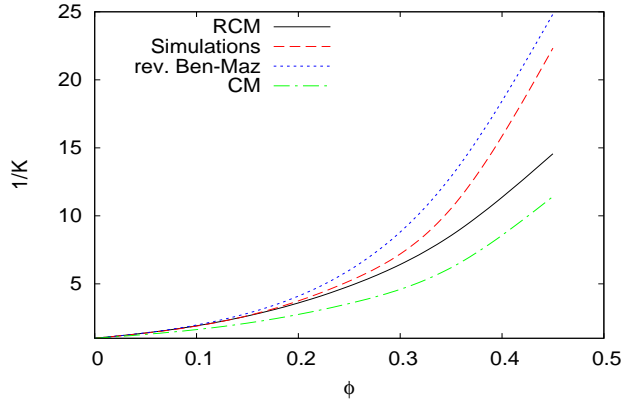}%
\caption{Inverse of the sedimentation coefficient $K$, equation (\ref{sedimentation}),
as a function of volume fraction $\phi$
for monodisperse suspension of hard-spheres in equilibrium.
Black (solid line) - the renormalized Clausius-Mossotti approximation introduced in this article,
red (long-dashed) line - numerical simulations of Abade at al. \cite{abade2010short},
blue (short-dashed) line - the revised Beenakker-Mazur method \cite{makuch2012transport},
green (dot-dashed) line - Clausius-Mossotti approximation defined by the equation (\ref{przbl na Tirr_c w CM}).}%
\label{rys sedymentacja}%
\end{figure}

\begin{figure}[ptbh]%
\centering
\includegraphics[
height=5.4982cm,
width=8.4987cm
]%
{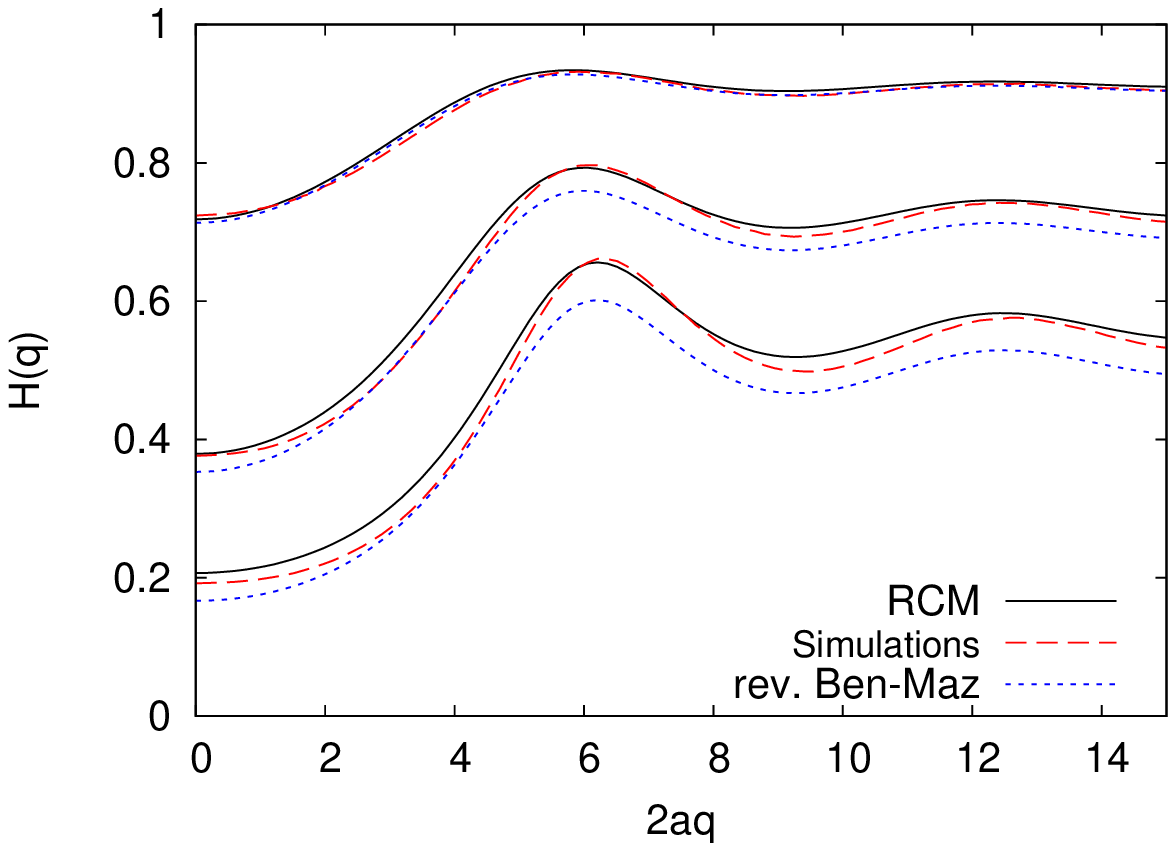}%
\caption{The hydrodynamic function $H\left(  q\right)  $, equation (\ref{hyd fun  mikro}), as a function of wave
vector for volume fractions $\phi=0.05,$ $\phi=0.15$ and $\phi=0.25$,
for monodisperse suspension of hard-spheres in equilibrium.
Black (solid line) - the renormalized Clausius-Mossotti approximation introduced in this article, red
(long-dashed) line - numerical simulations \cite{abade2010short}, blue
(short-dashed) line - the revised Beenakker-Mazur method \cite{makuch2012transport}.}%
\label{rys hydfun}%
\end{figure}

\begin{figure}[ptbh]%
\centering
\includegraphics[
height=5.4982cm,
width=8.4987cm
]%
{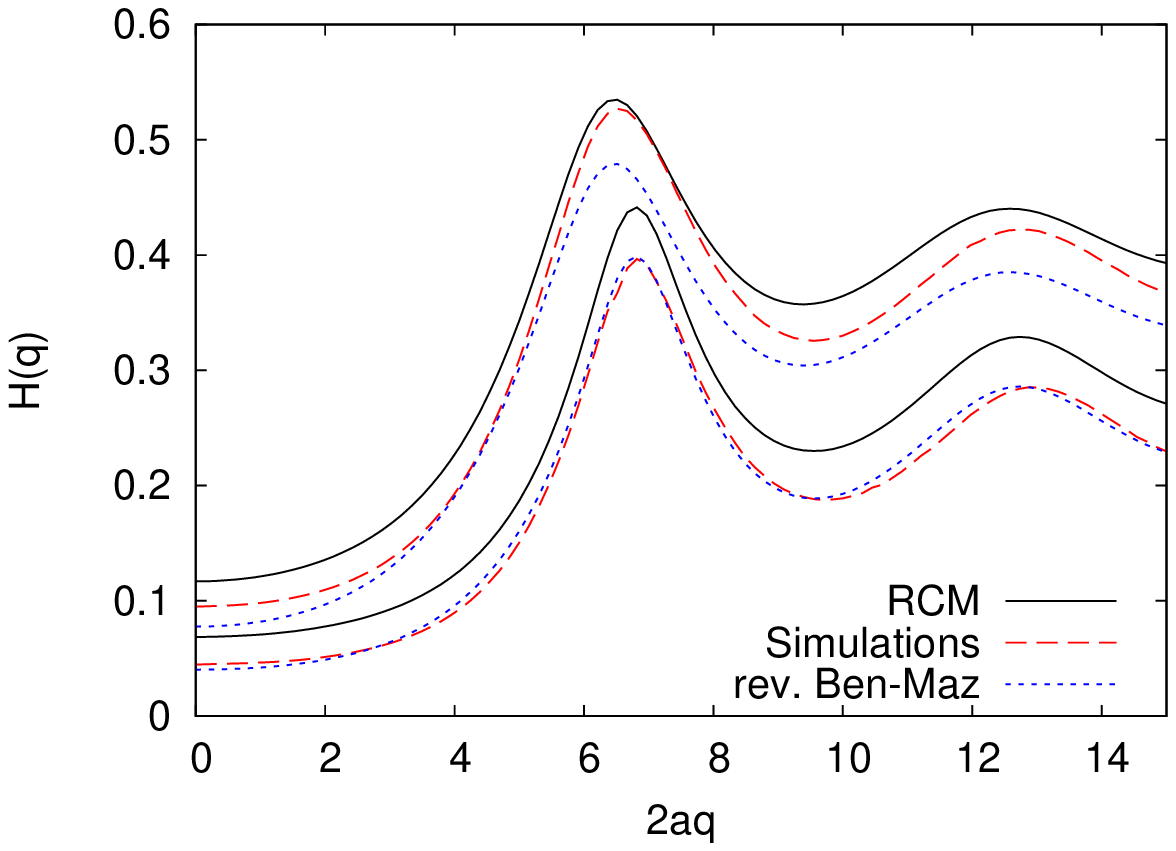}%
\caption{The hydrodynamic function $H\left(  q\right)  $, equation (\ref{hyd fun  mikro}), as a function of wave
vector for volume fractions $\phi=0.35$ and $\phi=0.45,$
for monodisperse suspension of hard-spheres in equilibrium.
Black (solid line) - the renormalized Clausius-Mossotti approximation introduced in this article, red
(long-dashed) line - numerical simulations \cite{abade2010short}, blue
(short-dashed) line - the revised Beenakker-Mazur method \cite{makuch2012transport}.}%
\label{rys hydfun 2}%
\end{figure}

\begin{figure}[ptbh]%
\centering
\includegraphics[
height=5.4982cm,
width=8.4987cm
]%
{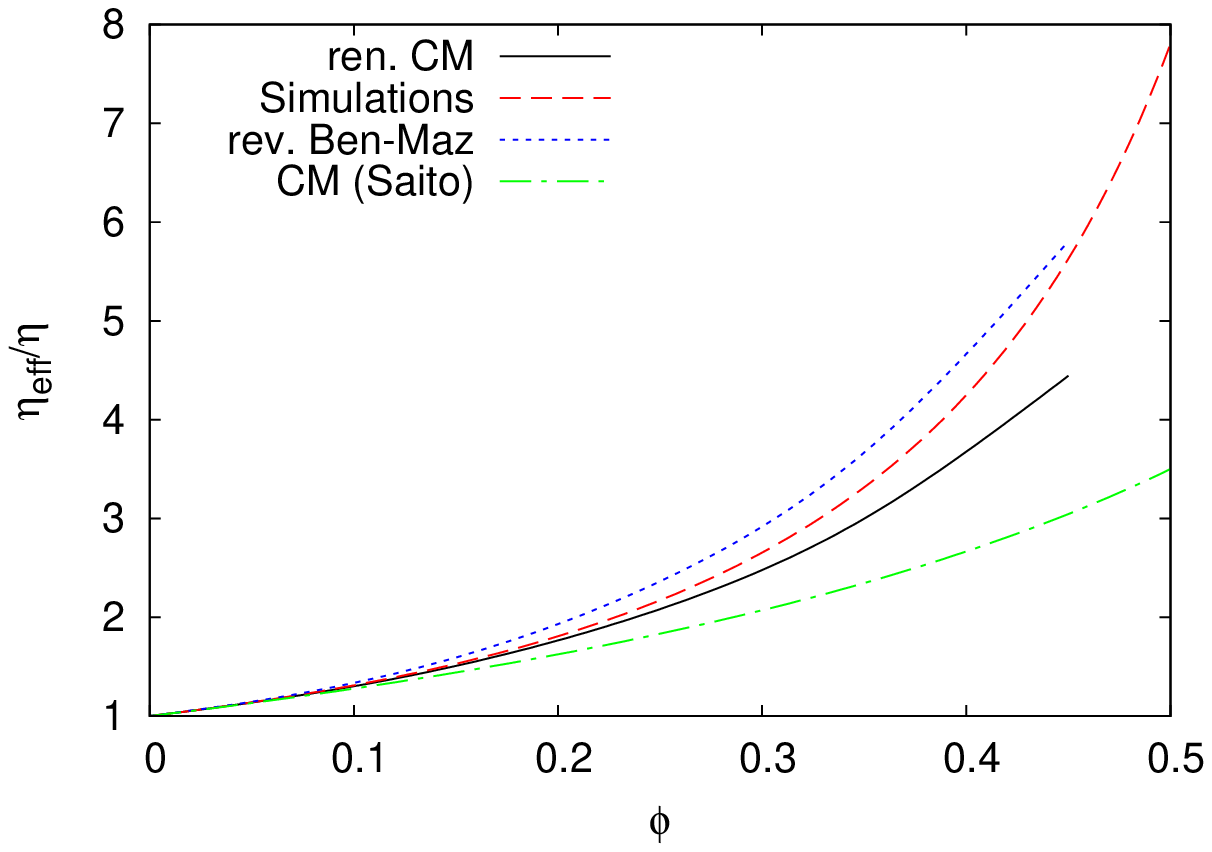}%
\caption{The relative effective viscosity $\eta_{\text{eff}}/\eta$ (high frequency, low shear), equation (\ref{eta micro 3}), 
as a function of volume fraction $\phi$
for monodisperse suspension of hard-spheres in equilibrium.
Black (solid line) - the renormalized Clausius-Mossotti approximation introduced in this article,
red (long-dashed) line - numerical simulations of Ladd \cite{JChemPhys_93_3484},
blue (short-dashed) line - the revised Beenakker-Mazur method \cite{makuch2012transport},
green (dot-dashed) line - Saito formula \cite{saito1950concentration}.}%
\label{rys lepkosc}%
\end{figure}

The renormalized Clausius-Mossotti approximation introduced in the previous
paragraph allows to calculate the short-time transport properties of
suspension, when the volume fraction $\phi$ and the two-body correlation
function $h\left(  \mathbf{R},\mathbf{R}^{\prime}\right)  $ are given. Within
the renormalized Clausius-Mossotti approximation, we perform calculations for
the volume fractions $\phi=0.05,0.15,0.25,0.35,$ and $0.45.$ For each volume
fraction, we use the two-particle correlation function in the Percus-Yevick
approximation for the hard-sphere potential \cite{smith2008fortran}.
Our results are presented in figs. \ref{rys self-diffusion}- \ref{rys lepkosc},
which show the translational short-time self-diffusion coefficient $D_{s}$,
the sedimentation coefficient $K$,
the hydrodynamic function $H(q)$, and the effective viscosity coefficient $\eta_{\text{eff}}$, respectively.

In the figs. \ref{rys self-diffusion}- \ref{rys lepkosc} we also present
results of the numerical simulations 
and the (revised) second order Beenakker-Mazur method
\cite{makuch2012transport}, which is nowadays the most
comprehensive theoretical scheme for calculations of the short-time transport
properties of suspensions.
At this point it is worth noting, that it is difficult to realize
experimentally a monodisperse suspension of hard-spheres
\cite{royall2013search} and to measure its volume fraction precisely
\cite{poon2012measuring}. However, if the experimental conditions satisfy the
assumptions underlying our theoretical model (such as monodisperse,
hard-sphere potential, regime of the zero Reynolds number), then this
suspension would have characteristics consistent with the precise numerical
simulations. Therefore, in this article, we assess the theoretical methods for
monodisperse hard-sphere suspensions by comparing them with numerical simulations,
instead of comparing with the experimental works
\cite{Segre1995,ottewill1987study,van1992long,Bergenholtz1998,van1989linear,Kruif1992high-frequency,Shikata1994,fritz2002high,PhysRevE.81.051402}.

Our results for the effective viscosity $\eta_{\text{eff}}$ and the
hydrodynamic function $H\left(  q\right)  $ (with its low and high $q$
behavior) presented in figs. \ref{rys self-diffusion}-\ref{rys lepkosc}
obtained within the renormalized Clausius-Mossotti approximation, when
compared with the numerical simulations and the revised second order
Beenakker-Mazur method - can be summarized as follows. For the volume
fractions $\phi\lesssim30\%,$ the relative error (with respect to the numerical
simulations) of the renormalized Clausius-Mossotii approximation is a few times
less or comparable with the relative error of the Beenakker-Mazur method -
it holds for the effective viscosity, the sedimentation coefficient, the
self-diffusion coefficient, and for almost whole range of the wave vectors $q$ of
the hydrodynamic function. The only exception is a range of the wave vectors
$2<2aq<5$ placed between $q=0$ and $q$ in the principal maximum of the
hydrodynamic function $H\left(  q\right)  $ (vide figs. \ref{rys hydfun} and
\ref{rys hydfun 2}). Here, the Beenakker-Mazur method is consistent with numerical simulations.
For volume fractions above $\phi \approx 30\%$, the Beenakker-Mazur method leads
to better agreement with numerical simulations
than the renormalized Clausius-Mossotti approximation,
for the effective viscosity and for most of the wave vectors $q$ of hydrodynamic function.

Before comparison of the results of our renormalized Clausius-Mossotti approximation defined by the equation (\ref{uog przybl CM})
with the Clausius-Mossotti approximation defined by the formula (\ref{przbl na Tirr_c w CM}),
it has to be emphasized that the term Clausius-Mossotti approximation may refer to two different variants of approximation.
First variant is given by the equations (\ref{przbl na Tirr_c w CM}) and (\ref{def operator CM}),
and is used in this article.
The second variant is also given by the equations (\ref{przbl na Tirr_c w CM}) and (\ref{def operator CM}),
but instead of the two-body correlation function $h$, its lowest virial term, i.e. the Mayer function for hard spheres \cite{hansen2006theory}, is used.
The Clausius-Mossotti approximation in case of the effective viscosity leads to the Saito formula (\ref{saito}),
whereas in case of the hydrodynamic function denoted in this approximation by $H_{CM}(q)$
- because of the fact that for the large wave vectors hydrodynamic function is related to the self-scattering sequences
and the Clausius-Mossotti approximation takes into account only one single particle-term
among self-scattering sequences
- it gives the following result
\begin{equation}
	\lim_{q \to \infty} H_{CM}(q) = 1.
\end{equation}
Therefore the self-diffusion coefficient in the Clausius-Mossotti approximation
does not depend on the volume fraction of suspension.
The opposite limit of the zero wave vector related to the sedimentation coefficient is presented in the fig. \ref{rys sedymentacja}.

It is worth shedding light on the results of our article - the derivation of the
ring expansion (\ref{Tirr rozw pierscieniowe}) and the formulation of the
renormalized Clausius-Mossotti approximation - from the perspective of the
hydrodynamic interactions and statistical physics.
The Beenakker-Mazur method
is currently the most comprehensive statistical physics method to calculate
the short-time transport properties of suspensions. With this article, we
introduce another method - the renormalized Clausius-Mossotti approximation.
Neither of the above approximations take the two-body hydrodynamic
interactions fully into account. It could be verified by a simple analysis of relevant equations on the two-body level.
Consequently, the strong
hydrodynamic interactions of close particles in suspensions are disregarded in
both approaches. Therefore, to construct a satisfactory method of calculations
of transport properties of suspensions, which would take the two-body
hydrodynamic interactions into consideration, remains an open problem of
statistical physics. It is worth noting here, that to take the two-body
hydrodynamic interactions fully into account in the Beenakker-Mazur
renormalized fluctuation expansion, a resummation up infinite order is needed.
An extension of the renormalized Clausius-Mossotti approximation to take the
two-body hydrodynamic interactions fully into consideration can be done, e.g.
by modification of the approximation (\ref{uog przybl CM}), adding the
two-body contributions.
This type of extension is natural,
because it goes along the line of a usual, systematic generalization of the Clausius-Mossotti approximation \cite{bedeaux1987effective}.
In order to fully grasp the two-body effect in the Beenakker-Mazur method,
one needs to consider all orders of the fluctuation expansion.
It is an important difference between the renormalized Clausius-Mossotti approximation and the Beenakker-Mazur method.
 
There is another intriguing point from the perspective of statistical physics.
The renormalized Clausius-Mossotti approximation
and the Beenakker-Mazur second order approach are similar,
because an important element of both methods is an effective propagator.
In the Beenakker-Mazur method, its role is played by the quantity
$\mathbf{A}_{\gamma_{0}},$ which depends on the volume fraction $\phi$, but
does not depend on the distribution of particles, e.g. the two-body
correlation function \cite{Beenakker1984Diffusion}. Therefore, this propagator
$\mathbf{A}_{\gamma_{0}}$ is the same for hard-sphere suspension and for
suspension of charged particles in equilibrium.
On the other hand,
the effective propagator $\mathbf{G}_{eff}$ in the ring expansion, on which
the renormalized Clausius-Mossotti approximation is constructed, depends both,
on the hydrodynamic interactions, and the distribution of particles.
This confrontation gives rise to the question concerning sensitivity of both methods to the
change in structure of suspension - when, for example,
an electrostatic interparticle repulsion increases in a suspension and the volume fraction remains unchanged.
This question in the case of the Beenakker-Mazur method has been answered in some situations:
the Beenakker-Mazur method is rather insensitive to the change in the structure of suspension
\cite{heinen2011short,Makuch2015rotational}.
We are going to address the above questions in our further work.

\section*{Acknowledgements}
I am very grateful to Professor Bogdan Cichocki,
who attracted my attention to the problem of transport properties of suspensions,
who inspired me to introduce the idea of the renormalization presented in this article
and who supervised me during my PhD study.
I also thank Maria Ekiel-Je\.zewska for her comments about the ring expansion.
The support of MNiSW grant IP2012 041572, and,
in the earlier part of the research work, support by the Foundation for Polish Science (FNP)
through the TEAM/2010-6/2 project, co-financed by the EU
European Regional Development Fund, is acknowledged.

\appendix

\section{Wave dependent sedimentation \label{app hq}}

The wave dependent sedimentation coefficient $H\left(  q\right)  $ describes
response of quiescent suspension, $\mathbf{v}_{0}=0$, cf. equations
(\ref{Stokes mobility integral forrm}), under action of a sinusoidal external
force density, e.g.%
\begin{equation}
\mathbf{f}_{\text{ext}}\left(  \mathbf{r}\right)  =f_{0}\mathbf{\hat{e}}%
_{z}\mathbf{\operatorname{Re}}\exp\left(  -iq\mathbf{\hat{e}}_{z}%
\cdot\mathbf{r}\right)  \exp\left(  \mathbf{-}\eta\left\vert z\right\vert
\right)  . \label{external force sedym}%
\end{equation}
The above external force density corresponds to the plane wave in the
direction of the wave vector $\mathbf{q}=q\mathbf{\hat{e}}_{z}$. For a moment,
we also introduce the damping factor $\exp\left(  -\eta\left\vert z\right\vert
\right)  $ and later we will take the limit $\eta\rightarrow0^{+}$. The above
external force is translationally invariant in $x$ and $y$ directions,
$\mathbf{f}_{\text{ext}}\left(  \mathbf{r}\right)  =\mathbf{\hat{e}}%
_{z}f_{\text{ext}}\left(  z\right)  .$ For homogeneous and isotropic
suspension that property induces the same form for the average velocity field,
i.e. $\left\langle \mathbf{v}\left(  \mathbf{r}\right)  \right\rangle
=\mathbf{\hat{e}}_{z}\left\langle v\left(  z\right)  \right\rangle $.
Incompressibility condition for the average velocity field $\left\langle
\mathbf{v}\left(  \mathbf{r}\right)  \right\rangle =\mathbf{\hat{e}}%
_{z}\left\langle v\left(  z\right)  \right\rangle $ implies $\left\langle
v\left(  z\right)  \right\rangle =$const. The force given by equation
(\ref{external force sedym}) has also the mirror symmetry in $z$ direction,
$f_{\text{ext}}\left(  z\right)  =-f_{\text{ext}}\left(  -z\right)  $, which,
for homogeneous and isotropic suspension, induces the same symmetry for the
velocity field $\left\langle v\left(  z\right)  \right\rangle =-\left\langle
v\left(  -z\right)  \right\rangle $. Along with the incompressibility
condition $\left\langle v\left(  z\right)  \right\rangle =$const, we obtain,
that the average velocity field vanishes in the whole suspension,
$\left\langle \mathbf{v}\left(  \mathbf{r}\right)  \right\rangle =0$.

This property of the zero net flux, $\left\langle \mathbf{v}\left(
\mathbf{r}\right)  \right\rangle =0$, simplify the equation (\ref{def Tirr}).
In this situation, the upper component of the vector $\left[  \left\langle
\mathbf{U}\right\rangle ,\left\langle \mathbf{f}\right\rangle \right]  $ in
the equation (\ref{def Tirr}) is given by%
\begin{equation}
\left\langle \mathbf{U}\left(  \mathbf{R,r}\right)  \right\rangle =\int
d^{3}R^{\mathbf{\prime}}d^{3}r^{\mathbf{\prime}}P_{U}\mathbf{T}^{\text{irr}%
}\left(  \mathbf{R,r};\mathbf{R}^{\prime},\mathbf{r}^{\prime}\right)
P_{U}^{T}\mathbf{f}_{\text{ext}}\left(  \mathbf{r}^{\prime}\right)  ,
\label{U and fext}%
\end{equation}
were projector $P_{U}$ (with its transposition $P_{U}^{T}$), by definition
projects on the upper half of the double vector $\left[  \left\langle
\mathbf{U}\right\rangle ,\left\langle \mathbf{f}\right\rangle \right]  ,$ as
follows from the formula (\ref{def Pu}). From now on, we will take the limit
$\eta\rightarrow0^{+},$ in which the zero net flux condition, $\left\langle
\mathbf{v}\left(  \mathbf{r}\right)  \right\rangle =0,$ remains.

The external force density given by equation (\ref{external force sedym}) is
torque-free, because the torque acting on the particle at position
$\mathbf{R}$ vanishes, $\int d^{3}r\theta\left(  \left\vert \mathbf{r}%
-\mathbf{R}\right\vert -a\right)  \left(  \mathbf{r-R}\right)  \times
\mathbf{f}_{\text{ext}}\left(  \mathbf{r}\right)  =0,$ for any position
$\mathbf{R.}$ The Heaviside function $\theta\left(  \left\vert \mathbf{r}%
-\mathbf{R}\right\vert -a\right)  $ used here, vanishes outside the particle
centered at $\mathbf{R}$ and is equal to $1$ inside the volume of the
particle. It is very simple to show, that in the case of the torque free
external force density, action of $\mathbf{M}_{>}$ operator given by equation
(\ref{single M>}), for such external force density for hard spheres, can be
written as%
\begin{equation}
\mathbf{f}_{i}\left(  \mathbf{R}_{1};\mathbf{r}\right)  =\frac{1}{\frac{4}%
{3}\pi a^{3}}\int d^{3}r^{\prime}\mathbf{M}_{>}\left(  \mathbf{R}%
_{1}\mathbf{,r,r}^{\prime}\right)  \mathbf{F}_{\text{ext}}\left(
\mathbf{R}_{1}\right)  , \label{M na gestosc sily bez momentu}%
\end{equation}
where $\mathbf{F}_{\text{ext}}\left(  \mathbf{R}\right)  $ is the total force
acting on a particle centered at position $\mathbf{R,}$%
\begin{equation}
\mathbf{F}_{\text{ext}}\left(  \mathbf{R}\right)  =\int d^{3}r\theta\left(
\left\vert \mathbf{r}-\mathbf{R}\right\vert -a\right)  \mathbf{f}_{\text{ext}%
}\left(  \mathbf{r}\right)  .
\end{equation}
The equation (\ref{M na gestosc sily bez momentu}) is a straightforward
consequence of the form of $\mathbf{M}_{>}$ given in the reference
\cite{makuch2012scattering}. Similar holds for the single particle
$\mathbf{M}_{0}$ operator,%
\begin{equation}
\mathbf{U}_{1}\left(  \mathbf{R}_{1};\mathbf{r}\right)  =\frac{1}{\frac{4}%
{3}\pi a^{3}}\int d^3r^{\prime}\mathbf{M}_{0}\left(  \mathbf{R}%
_{1}\mathbf{,r,r}^{\prime}\right)  \mathbf{F}_{\text{ext}}\left(
\mathbf{R}_{1}\right)  . \label{M0 na gestosc sily bey momentu}%
\end{equation}
Therefore, action of the operators $\mathbf{M}_{>}$ and $\mathbf{M}_{0}$ on
the force density $\mathbf{f}_{\text{ext}}$ in equations
(\ref{Stokes mobility integral form}), can be replaced using
(\ref{M na gestosc sily bez momentu}) and
(\ref{M0 na gestosc sily bey momentu}). It has the following implication in
the equation (\ref{U and fext}),%
\begin{equation}
\left\langle \mathbf{U}\left(  \mathbf{R,r}\right)  \right\rangle =\frac
{1}{\frac{4}{3}\pi a^{3}}\int d^{3}R^{\mathbf{\prime}}\int d^{3}%
r^{\mathbf{\prime}}P_{U}\mathbf{T}^{\text{irr}}\left(  \mathbf{R,r}%
;\mathbf{R}^{\prime},\mathbf{r}^{\prime}\right)  P_{U}^{T}\mathbf{F}%
_{\text{ext}}\left(  \mathbf{R}^{\prime}\right)  . \label{U and Fext}%
\end{equation}
In the problem of sedimentation, we consider the average translational
velocity of particles, $\left\langle \mathbf{V}\left(  \mathbf{R}\right)
\right\rangle =\left\langle \sum_{i=1}^{N}\delta\left(  \mathbf{R-R}%
_{i}\right)  \mathbf{V}_{i}\left(  X\right)  \right\rangle $. For hard
spheres, $\left\langle \mathbf{V}\left(  \mathbf{R}\right)  \right\rangle $ is
related to the average velocity field of particles $\left\langle
\mathbf{U}\left(  \mathbf{R,r}\right)  \right\rangle ,$ by the formula%
\begin{equation}
\left\langle \mathbf{V}\left(  \mathbf{R}\right)  \right\rangle =\frac
{1}{\frac{4}{3}\pi a^{3}}\int d^{3}r\left\langle \mathbf{U}\left(
\mathbf{R,r}\right)  \right\rangle ,
\end{equation}
which follows from the relation (\ref{def particle field}).

The expression (\ref{U and Fext}), after passing from the particle velocity
field $\left\langle \mathbf{U}\left(  \mathbf{R,r}\right)  \right\rangle $ for
hard spheres, to the average translation velocity $\left\langle \mathbf{V}%
\left(  \mathbf{R}\right)  \right\rangle ,$ yields%
\begin{equation}
\left\langle \mathbf{V}\left(  \mathbf{R}\right)  \right\rangle =\int
d^{3}R^{\mathbf{\prime}}\mathbf{Y}\left(  \mathbf{R,R}^{\prime}\right)
\mathbf{F}_{\text{ext}}\left(  \mathbf{R}^{\prime}\right)  ,
\label{app V Fext}%
\end{equation}
where%
\begin{equation}
\mathbf{Y}\left(  \mathbf{R,R}^{\prime}\right)  =\frac{1}{\left(  \frac{4}%
{3}\pi a^{3}\right)  ^{2}}\int d^{3}r\int d^{3}r^{\mathbf{\prime}}%
P_{U}\mathbf{T}^{\text{irr}}\left(  \mathbf{R,r};\mathbf{R}^{\prime
},\mathbf{r}^{\prime}\right)  P_{U}^{T}.
\end{equation}
For homogeneous suspension, the above kernel $\mathbf{Y}\left(  \mathbf{R,R}%
^{\prime}\right)  $ depends only on the difference of positions, therefore we
can introduce%
\begin{equation}
\mathbf{Y}\left(  \mathbf{R-R}^{\prime}\right)  \equiv\mathbf{Y}\left(
\mathbf{R,R}^{\prime}\right)  .
\end{equation}
Moreover, isotropy implies, that $\mathbf{Y}\left(  \mathbf{R}\right)  $ is of
the form $y_{0}\left(  R\right)  \mathbf{1+}y_{2}\left(  R\right)
\mathbf{\hat{R}\hat{R}}$. This form of $\mathbf{Y}\left(  \mathbf{R}\right)
,$ along with simple algebraic manipulations of equations (\ref{app V Fext}),
leads to the following conclusion. For the external force given by a plane
wave,%
\begin{equation}
\mathbf{F}_{\text{ext}}\left(  \mathbf{R}\right)  =F_{0}\mathbf{\hat
{q}\operatorname{Re}}\exp\left(  -i\mathbf{q}\cdot\mathbf{R}\right)  ,
\end{equation}
the average translational velocity has also the plane-wave form,%
\begin{equation}
\left\langle \mathbf{V}\left(  \mathbf{R}\right)  \right\rangle =V\left(
q\right)  \mathbf{\hat{q}\operatorname{Re}}\exp\left(  -i\mathbf{qR}\right)  ,
\end{equation}
where the coefficient $V\left(  q\right)  $ is given by the formula%
\[
V\left(  q\right)  =H\left(  q\right)  \mu_{0}F_{0},
\]
with hydrodynamic function given by equation (\ref{hyd fun mikro}). The stokes
coefficient $\mu_{0}=1/(6\pi\eta a)$.

\section{Calculations within renormalized Clausius-Mossotti approximation
\label{app solution}}

In this appendix we give some details of our calculations within the
renormalized Clausius-Mossotti approximation. It demands to solve the set of
equations (\ref{uog przybl CM}), (\ref{def ren operator CM}),
(\ref{Tirr and T}), (\ref{Geff}), (\ref{T into self off}) and
(\ref{self by off}) for the quantities $\mathbf{B,}$ $\mathbf{T}_{RCM}^{irr}$,
$\mathbf{G}_{eff},$ $\mathbf{T}^{irr}$, $\mathbf{T,}$ $\mathbf{T}_{\text{off}}$. 
Each of those quantities is a fourfold $3 \times 3$ matrix, as it may be inferred for example
from the formula (\ref{def T}) for the $\mathbf{T}$ operator.
Therefore the $\mathbf{T}$ operator written with all variables and indexes is represented by
$\mathbf{T}(\mathbf{R,r,R',r'})_{u \alpha u^{\prime}, \alpha^{\prime}}$.
It has two indexes $u,u^{\prime}=U,P$, which denotes upper or lower part of the double vector,
and another two Cartesian indexes $\alpha,\alpha^{\prime}=1,2,3$.
In our calculations we represent those quantities as multipole hydrodynamic matrices \cite{cichocki1988hydrodynamic}.
In the reference \cite{makuch2012scattering} the reader can find how to introduce a multipole picture
for the forces and velocities and how to represent the hydrodynamic matrices $\mathbf{M}$ and $\mathbf{G}$ in the multipole picture.
The notation used in the reference \cite{makuch2012scattering} is adopted also in this article,
therefore we do not repeat that material.
In the same way as in the reference \cite{makuch2012scattering} we introduce the multipole picture of the hydrodynamic matrices 
$\mathbf{B,}$ $\mathbf{T}_{RCM}^{irr}$, $\mathbf{G}_{eff},$ $\mathbf{T}^{irr}$, $\mathbf{T}$ and $\mathbf{T}_{\text{off}}$.
In the multipole picture, all of the above quantities become infinite dimensional
hydrodynamic multipole matrices, e.g. $\mathbf{T}\left(  \mathbf{R}%
,\mathbf{r};\mathbf{R}^{\prime},\mathbf{r}^{\prime}\right)_{u \alpha u^{\prime}, \alpha^{\prime}}  \rightarrow$
$\left[  T\left(  \mathbf{R},\mathbf{R}^{\prime}\right)  \right]
_{ulm\sigma,u^{\prime}l^{\prime}m^{\prime}\sigma^{\prime}}$.
Therefore the variables $\mathbf{r},\mathbf{r}^{\prime}$ and Cartesian indexes are
transformed into multipole numbers $l,m$ and $\sigma$ having the following range:
$l=1,2,\ldots,\infty,$ $m=-l,-l+1,\ldots,l,$ whereas $\sigma=0,1,2$.
An important role in our calculations is played by the homogeneity of the system,
because it implies that the matrices depend on the difference of positions only,
for example 
$\left[ T\left(  \mathbf{R}-\mathbf{R}^{\prime}\right)  \right] _{ulm\sigma,u^{\prime}l^{\prime}m^{\prime}\sigma^{\prime}}
\equiv \left[ T\left(  \mathbf{R},\mathbf{R}^{\prime}\right)  \right] _{ulm\sigma,u^{\prime}l^{\prime}m^{\prime}\sigma^{\prime}}$.
We also use isotropy of the system, which allow to calculate the multipole matrix
$T\left(  \mathbf{R} \right)$ for any vector $\mathbf{R}$ when the $T$ for $z$ direction $\mathbf{R}=R\mathbf{e}_z$ is known.

Using also the Fourier space, with the Fourier transformation given by 
\begin{equation}
\hat{T}(\mathbf{k})=\int d^3R \ exp[-i\mathbf{k}\mathbf{R}] T(\mathbf{R}),
\label{def Fourier}
\end{equation}
with similar definition for other multipole hydrodynamic matrices,
we represent the set of equations 
(\ref{uog przybl CM}), (\ref{def ren operator CM}), (\ref{Tirr and T}), (\ref{Geff}), (\ref{T into self off}) and (\ref{self by off})
as follows
\begin{equation}
\hat{T}(\mathbf{k})=\hat{B}+\hat{T}_{\text{off}}(\mathbf{k}), \label{T into self off multip}%
\end{equation}
\begin{equation}
\hat{T}(\mathbf{k})=\hat{T}^{\text{irr}}(\mathbf{k})\left(  1 - \hat{G}(\mathbf{k}) \hat{T}^{\text{irr}}(\mathbf{k}) \right)^{-1},
\end{equation}
\begin{equation}
\hat{G}_{\text{eff}}(\mathbf{k})=\hat{G}(\mathbf{k})+\hat{G}(\mathbf{k})\hat{T}(\mathbf{k})\hat{G}(\mathbf{k}), \label{Geff multip}%
\end{equation}
\begin{equation}
\hat{T}_{\text{RCM}}^{\text{irr}}(\mathbf{k})\approx\hat{B}, \label{uog przybl CM multip}%
\end{equation}
\begin{equation}
\hat{T}^{\text{irr}}(\mathbf{k})=\hat{T}_{\text{RCM}}^{\text{irr}}(\mathbf{k})\left(  
1-\widehat{\left[ h{G}_{\text{eff}} \right]}(\mathbf{k})  \hat{T}_{\text{RCM}}^{\text{irr}}(\mathbf{k})\right)  ^{-1},
\label{def ren operator CM multip}%
\end{equation}
\begin{equation}
\hat{B} =n_{1}M  + \int d^3R \ T_{\text{off}}(-\mathbf{R})G(\mathbf{R})M,
\label{self by off multip}%
\end{equation}
where
\begin{equation}
\widehat{\left[ h{G}_{\text{eff}} \right]}(\mathbf{k})=\int d^3R \ exp[-i\mathbf{k}\mathbf{R}] h(\mathbf{R}){G}_{\text{eff}}(\mathbf{R}).
\end{equation}
In the above equations there appear superpositions of the multipole hydrodynamic matrices,
inversion of matrices, and the matrices appear both in the positional and the Fourier space.
We created numerical code to solve these equations of the renormalized
Clausius-Mossotti approximation. Three aspects appear here.  

\begin{figure}[ptbh]%
\centering
\includegraphics[
height=5.4982cm,
width=8.4987cm
]%
{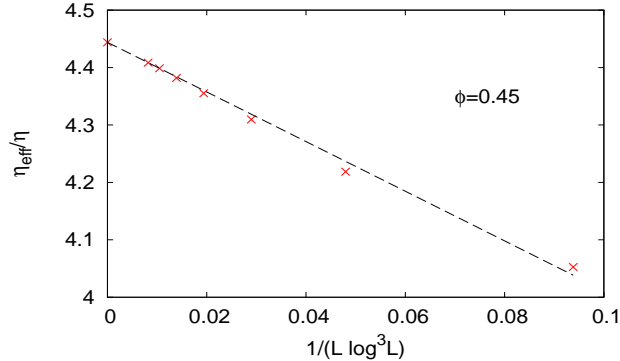}%
\caption{The relative effective viscosity coefficient $\eta_{eff}/\eta$ as a
function of multipole truncation $L$ obtained by the renormalized
Clausius-Mossotti approximation for suspension of volume fraction $\phi
=0.45$.}%
\label{rys viscosity L truncation}%
\end{figure}

First, in our numerical calculations, we truncate the hydrodynamic matrices,
e.g.$\left[  \hat{T}\left(  \mathbf{k} \right)  \right]
_{ulm\sigma,u^{\prime}l^{\prime}m^{\prime}\sigma^{\prime}}$.
The truncation is characterized by $L,$ which is the
highest multipole used in the calculations - we consider matrix elements with
$l,l^{\prime}\leq L$ only. The calculations were performed for different
parameters truncation, $L=4,\ldots,10$. The dependence of the effective viscosity
coefficient $\eta_{\text{eff}}/\eta$ on the function $1/\left(  L\log
^{3}L\right)  $ of the truncation parameter $L$ is presented in the fig.
\ref{rys viscosity L truncation}. The figure shows, that even for the highest
truncation parameter $L=10$ (which corresponds to $1/\left(  L\log
^{3}L\right)  \approx0.008$) the effective viscosity coefficient is still sensitive
to the truncation $L$. Therefore, we extrapolate the coefficient
up to $L\rightarrow\infty$. The extrapolated value of the
coefficient is given by intersection of a straight line,
passing through the points corresponding to $L=9$ and $L=10$ in the fig.
\ref{rys viscosity L truncation}, with the vertical axis. Similar procedure of
the extrapolation is carried in the case of the other transport characteristics.

The second numerical aspect is related to a discretization of distance $R$ for
hydrodynamic matrices, e.g. $T\left(  R\mathbf{e}_z\right)  $. Points $R=\xi_{0}%
,\ldots,\xi_{N},$ with $\xi_{0}=0$ and $\xi_{n}=\xi_{1}\exp\left[
\alpha\left(  i-1\right)  \right]  $ were considered. The parameter $\alpha$ was
determined from the assumption, that the first and the last section are equal,
i.e., $\xi_{1}=\xi_{N}-\xi_{N-1}$. Therefore, $\xi_{1}$ and $N$ determine a
set of points, in which the hydrodynamic matrices were considered in the code.
Other values - if needed - were calculated by interpolation or from the
asymptotic expansion (e.g. for $R$ larger than $\xi_{N}$).
It is worth noting that the exponential mesh is convenient to calculate the three
dimensional Fourier transform of the hydrodynamic matrices.
In our calculations the three dimensional Fourier transform of hydrodynamic matrices
in the formula (\ref{def Fourier}) was first reduced to the one dimensional 
Hankel transform \cite{bleistein1975asymptotic} in a similarity to a dielectric system \cite{makuch2015multipole}.
Then the exponential mesh is used to perform Hankel transform with the use of
numerical procedures for fast Fourier transform.
We performed calculations for $\xi_{1}/\left(  2a\right)  =1/2,1/3,1/4,1/5$ and
$N=512,1024,2048,4096$ respectively. Larger $N$ correspond to a denser mesh.
We found the mesh characterized by $N=4096$ sufficient and it is used to
obtain the results presented in this article.


The third aspect of the numerical calculations is related to the fact, that
the system of equations was solved iteratively. We observed, that up to the
volume fractions $\phi\approx45\%,$ after a few iterations, difference between
hydrodynamic functions in subsequent iterations decays as it happens in a geometric
series. It may be written as follows%
\begin{equation}
\sup_{q}\left\vert H_{i}\left(  q\right)  -H_{i-1}\left(  q\right)
\right\vert \approx\sup_{q}\left\vert H_{i+1}\left(  q\right)  -H_{i}\left(
q\right)  \right\vert \Delta,
\end{equation}
where $H_{i}\left(  q\right)  $ denotes the hydrodynamic function after the
$i$-th iteration and symbol $\sup_{q}f\left(  q\right)  $ stands for maximal
value of a function $f$. The highest observed value of $\Delta\approx0.7$. The
iteration procedure was stopped when the following condition $\sup
_{q}\left\vert H_{i}\left(  q\right)  -H_{i-1}\left(  q\right)  \right\vert
<10^{-4}$ was satisfied.

It is worth mentioning, that the computer time and memory to solve the
equations iteratively is comparable with the calculations, which we performed in
the case of the revised Beenakker-Mazur method \cite{makuch2012scattering}.
The numerical results within the renormalized Clausius-Mossotti approximation
presented in this article were calculated
with a use of a desktop computer within one day.%

%

\end{document}